\documentclass[longauth]{aaEC}

\usepackage{graphicx}
\usepackage{natbib}
\usepackage{scalerel}

\usepackage[table]{xcolor}

\usepackage{txfonts}
\usepackage[pdfencoding=auto,psdextra]{hyperref}
\hypersetup{
    colorlinks=true,
    linkcolor=blue,
    filecolor=magenta,      
    urlcolor=blue,
    citecolor=blue
}
\makeatletter
\renewcommand*\aa@pageof{, page \thepage{} of \pageref*{LastPage}}
\makeatother

\usepackage[utf8]{inputenc}

\usepackage[switch, modulo]{lineno}
\usepackage{euclid}
\usepackage{physics}
 \usepackage{booktabs}

\begin{document}

\title{KiDS-Legacy: Joint analysis of second- and third-order cosmic shear}

\newcommand{\orcid}[1]{} %% define as link to https://orcid.org/#1 if needed
\author{\normalsize L.~Linke$^{1}$\thanks{\email{laila.linke@uibk.ac.at}}, 
L.~Porth$^{2}$, 
P.~Burger$^{2,3,4}$, 
J.~Harnois-Déraps$^{5}$,
S.~Heydenreich$^{2, 6}$, 
P.~Schneider$^{2}$, 
M. Asgari$^{5}$,
M. Bilicki$^{7}$,
C. Georgiou$^{8}$,
C. Heymans$^{9,10}$,
H. Hildebrandt$^{10}$,
H. Hoekstra$^{11}$,
P. Jalan$^{22}$,
B. Joachimi$^{12}$,
S. Joudaki$^{13, 14}$
K. Kuijken$^{11}$,
S. Li$^{15, 16}$,
L. Moscardini$^{17, 18, 19}$,
M. Radovich$^{20}$,
R. Reischke$^{2,10}$,
B. Stölzner$^{10}$,
A. H. Wright$^{10}$,
Z. Yan$^{21}$,
Y.-H. Zhang$^{9,11}$
}

\institute{$^{1}$ Universität Innsbruck, Institut für Astro- und Teilchenphysik, Technikerstr. 25/8, 6020 Innsbruck, Austria\\
$^{2}$ Universität Bonn, Argelander-Institut für Astronomie, Auf dem
Hügel 71, 53121 Bonn, Germany \\
$^{3}$ Waterloo Centre for Astrophysics, University of Waterloo, Waterloo,
Ontario N2L 3G1, Canada\\
$^{4}$ Department of Physics and Astronomy, University of Waterloo, Waterloo, ON N2L 3G1, Canada\\
$^{5}$ School of Mathematics, Statistics and Physics, Newcastle University, Newcastle-upon-Tyne, NE1 7RU, UK \\
$^{6}$ Department of Astronomy and Astrophysics, University of California, Santa Cruz, 1156 High Street, Santa Cruz, CA 95064, USA\\
$^{7}$ Center for Theoretical Physics, Polish Academy of Sciences, al. Lotników 32/46, 02-668 Warsaw, Poland\\
$^{8}$ Institut de Física d’Altes Energies (IFAE), The Barcelona Institute of Science and Technology, Campus UAB, 08193 Bellaterra (Barcelona), Spain\\
$^{9}$ Institute for Astronomy, University of Edinburgh, Royal Observatory, Blackford Hill, Edinburgh, EH9 3HJ, UK\\ 
$^{10}$ Ruhr University Bochum, Faculty of Physics and Astronomy, Astronomical Institute (AIRUB), German Centre for Cosmological Lensing, 44780 Bochum, Germany\\
$^{11}$ Leiden Observatory, Leiden University, P.O.Box 9513, 2300RA Leiden, The Netherlands\\
$^{12}$ Department of Physics and Astronomy, University College London, Gower Street, London WC1E 6BT, UK
$^{13}$ Centro de Investigaciones Energéticas, Medioambientales y Tecnológicas (CIEMAT), Av. Computense 40, E-28040 Madrid, Spain\\
$^{14}$ Institute of Cosmology \& Gravitation, Dennis Sciama Building, University of Portsmouth, Portsmouth, PO1 3FX, United Kingdom\\
$^{15}$ Kavli Institute for Particle Astrophysics and Cosmology, Stanford University, Stanford, CA 94305, USA\\
$^{16}$ SLAC National Accelerator Laboratory, Menlo Park, CA 94025, USA\\
$^{17}$ Dipartimento di Fisica e Astronomia "Augusto Righi" - Alma Mater Studiorum Università di Bologna, via Piero Gobetti 93/2, I-40129 Bologna, Italy\\
$^{18}$ Istituto Nazionale di Astrofisica (INAF), Osservatorio di Astrofisica e Scienza dello Spazio (OAS), via Piero Gobetti 93/3, I-40129 Bologna, Italy \\
$^{19}$ Istituto Nazionale di Fisica Nucleare (INFN), Sezione di Bologna, viale Berti Pichat 6/2, I-40127 Bologna, Italy\\
$^{20}$ INAF - Osservatorio Astronomico di Padova, via dell'Osservatorio 5, 35122 Padova, Italy\\
$^{21}$ Kobayashi-Maskawa Institute for the Origin of Particles and the Universe (KMI), Nagoya University, Nagoya, 464-8602, Japan\\
$^{22}$ Institute of Astronomy, National Tsing Hua University, 101, Section 2, Kuang-Fu Road, Hsinchu, 30013, Taiwan
}

\abstract{
Weak %gravitational 
lensing by large-scale structure is a powerful cosmological probe. While most %current 
analyses rely on second-order %shear 
correlations, these are primarily sensitive to the parameter combination $S_8 = \sigma_8 (\Omega_\mathrm{m}/0.3)^{0.5}$, limiting their ability to constrain $\Omega_\mathrm{m}$ and other cosmological parameters independently. Higher-order statistics capture non-Gaussian features of the density field and can therefore break parameter degeneracies and extract more cosmological information from weak lensing surveys. 
    }
    {
We present a joint analysis of second- and third-order cosmic shear in the final data release of the Kilo-Degree Survey (KiDS-Legacy). Our aim is to test whether the inclusion of non-Gaussian information improves cosmological constraints. %and provides new insights into the origin of the apparent discrepancy between late-time weak lensing and early-Universe CMB probes. 
    }
    {
    We combine COSEBIs (Complete Orthogonal Sets of E-/$B$-mode Integrals)  at scales between \ang{;2;} and \ang{;300;} with third-order aperture mass moments at scales between \ang{;4;} and \ang{;32;} to perform a joint analysis of second- and third-order %summary 
    statistics. Compared to previous KiDS analyses, we implement several methodological advances: an intrinsic alignment model with redshift and mass dependence, a baryon correction model validated on multiple hydrodynamical simulations, and corrections for reduced shear and source clustering. We further develop neural-network emulators to accelerate the inference.%, enabling efficient Bayesian parameter estimation. A strict blinding procedure ensures robustness against confirmation bias. 
    }
{
Combining COSEBIs with third-order aperture mass statistics in KiDS-Legacy yields $\Omega_\mathrm{m} = 0.297^{+0.056}_{-0.040}$ and $S_8 = 0.806^{+0.025}_{-0.023}$, significantly tightening the $\Omega_\mathrm{m}$ constraints and more than doubling the figure of merit in the $\Omega_\mathrm{m}$–$S_8$ plane compared to the two-point analysis alone. The third-order measurements pass stringent internal consistency tests %under catalogue-level splits 
and are fully compatible with the KiDS-Legacy two-point constraints.

}
{
The KiDS-Legacy joint second- and third-order analysis delivers competitive cosmological constraints that are consistent with other 2+3-point weak lensing results and with \textit{Planck} CMB measurements within $1\sigma$, providing no evidence for an $S_8$ tension %in this improved KiDS data set 
and demonstrating the maturity of third-order cosmic shear as a key probe for forthcoming surveys such as \textit{Euclid}. % and LSST.
}

    \keywords{cosmology: observations -- gravitational lensing: weak -- large-scale structure of Universe -- cosmological parameters}

   \titlerunning{KiDS-Legacy: 2+3-point cosmic shear}
   \authorrunning{Linke et al.}
   
   \maketitle
  
\section{\label{sc:Intro}Introduction}
Cosmology with cosmic shear, the correlation of galaxy shape distortions by the cosmic large-scale structure, is entering an interesting period. The first cosmological data of the next generation surveys like \Euclid \citep{EuclidSkyOverview} and LSST \citep{Ivezic2019} are imminent, while current surveys like the Dark Energy Survey (DES, \citealp{Abbott2016, Becker2016}), the Hyper-Supreme Cam survey (HSC, \citealp{Aihara2018}), and the Kilo-Degree Survey (KiDS, \citealp{Kuijken2015}) have finalized their fiducial cosmic shear analyses \citep{DESY6cosmicshear, Dalal2023, stoelzner2025}. These final data sets provide their tightest cosmological constraints and benefit from a decade of detailed understanding of each survey's characteristics.

Cosmic shear analyses from KiDS, DES, and HSC consistently show that this probe mostly constraints the parameter $S_8=\sigma_8 \sqrt{\Omm/0.3}$; a combination of the matter density parameter $\Omm$ and the clustering parameter $\sigma_8$. Strikingly, most cosmic shear measurements yield values for $S_8$ that are lower than the value inferred from CMB observations by the Planck satellite \citep{PlanckCosmologicalParameters}, sometimes by several standard deviations. This apparent mismatch between early- and late-time probes of the Universe has been dubbed the `$S_8$-tension' \citep{DiValentino2021}. 

The standard analysis of cosmic shear employs second-order statistics of observed galaxy shapes. These probe the cosmological information content in the Gaussian part of the cosmic large-scale structure. However, non-linear structure growth generates non-Gaussian features that two-point functions cannot capture. To exploit this additional information, a range of higher-order statistics (HOS) has been developed (see \citealp{Ajani-EP29} for a discussion of ten of them).

Among them, third-order correlations of the shear field are particularly promising. They extend naturally beyond two-point functions by probing the skewness, rather than the variance, of the shear field. Unlike many other HOS, third-order shear correlations can be modelled analytically without relying exclusively on large suites of simulations \citep{Heydenreich2023}. Models for astrophysical systematics such as intrinsic alignments (IA) and baryonic feedback have also been generalised from two- to three-point cosmic shear \citep{Pyne2021, Linke2024, Burger2025}. Although measuring higher-order statistics is generally more demanding, efficient algorithms \citep[e.g.][]{Porth2024} now make third-order shear measurements feasible for the wide-area data sets of upcoming surveys like \Euclid. Recent applications to current surveys \citep[e.g.][]{Burger2024,Gomes2025,Sugiyama2025} have demonstrated that robust third-order shear analyses are possible. Importantly, third-order cosmic shear probes a different combination of $\sigma_8$ and $\Omega_\mathrm{m}$ than two-point statistics. Their joint analysis therefore breaks the degeneracy between these parameters, substantially improves constraints in the $\sigma_8$–$\Omega_\mathrm{m}$ plane, and provides a means to disentangle whether the `$S_8$ tension’ originates from $\sigma_8$ or from $\Omega_\mathrm{m}$.

In this work we combine two- and three-point cosmic shear measured in the final data release of KiDS, dubbed KiDS-Legacy \citep{Wright2024}. This data set provides the most internally consistent KiDS catalogue %to date 
\citep{stoelzner2025}, incorporating several methodological improvements over previous releases. It covers a larger area, includes additional photometric measurements that improve photometric redshift estimates, and implements updated calibration methods compared to the previous data release, KiDS-1000 \citep{Kuijken2019, Wright2020, Hildebrandt2021, Giblin2021}. The $S_8$ value inferred from the KiDS-Legacy two-point analysis is more consistent with \textit{Planck} than earlier KiDS results. By analysing third-order statistics from the same catalogue, we aim to confirm or challenge this reduction of the `$S_8$ tension’.

Compared to the second- and third-order analysis of the previous KiDS data release by \citet{Burger2024}, we introduce several modelling advances designed to increase robustness: a refined baryon correction model \citep{Burger2025}, an IA model with explicit dependencies on stellar mass and redshift, and the inclusion of reduced shear and source clustering effects.

This paper is structured as follows. In Sect.~\ref{sc:Model} we describe our modelling of two- and three-point statistics, with emphasis on the changes relative to \citet{Burger2024}. Section \ref{sc:Data} introduces the simulated data sets and the KiDS-Legacy observations used in this analysis. Section~\ref{sc:Methods} details our measurement pipeline, blinding procedure, covariance construction, and parameter inference. Section~\ref{sc:Results} presents our results %on simulated and observed data
, and we conclude in Sect.~\ref{sc:Discussion} with a discussion of their implications.

\section{\label{sc:Model} Modelling of second- and third-order cosmic shear statistics}

\subsection{COSEBIs}
We adopt COSEBIs (Complete Orthogonal Sets of E-/$B$-mode Integrals; \citealt{Schneider2010,Asgari2012}) as our second-order summary statistic, following their use in the KiDS-Legacy cosmic shear analysis \citep{Wright2025}. Compared to the shear correlation functions $\xi_\pm(\theta)$, COSEBIs offer several advantages: they allow an exact separation of E- and $B$-modes on finite angular ranges, they compress cosmological information into a small number of modes, and they provide robust null tests through the $B_n$. Since weak lensing in standard gravity produces only E-modes, any significant detection of $B_n$ indicates residual systematics or non-standard physics. This property has already proven useful in KiDS, where spurious $B$-modes in an early version of the Legacy data set revealed an error in the astrometry, which was subsequently accounted for by masking affected pointings \citep{Wright2025}.

The shear correlation functions are defined as
\begin{equation}
    \xipmT{ij}(\theta) = \langle \gamt^{(i)}(\boldsymbol{\vartheta}) \gamt^{(j)}(\boldsymbol{\vartheta}+\boldsymbol{\theta}) \rangle 
    \pm \langle \gamx^{(i)}(\boldsymbol{\vartheta}) \gamx^{(j)}(\boldsymbol{\vartheta}+\boldsymbol{\theta}) \rangle ,
\end{equation}
where $\gamt^{(i)}$ and $\gamx^{(i)}$ are the tangential and cross shear components of tomographic bin $i$.  
The COSEBI modes are then obtained by projecting $\xipm$ with a set of orthogonal filter functions $\Tcosebi{n\pm}(\theta)$,
\begin{align}
    \notag \EnT{n}{ij} &= \frac{1}{2} \int_{\theta_\mathrm{min}}^{\theta_\mathrm{max}} \mathrm{d}\theta \, \theta 
    \left[ \Tcosebi{n+}(\theta) \, \xipT{ij}(\theta) + \Tcosebi{n\-}(\theta) \, \ximT{ij}(\theta) \right], \\
    \BnT{n}{ij} &= \frac{1}{2} \int_{\theta_\mathrm{min}}^{\theta_\mathrm{max}} \mathrm{d}\theta \, \theta 
    \left[\Tcosebi{n+}(\theta) \, \xipT{ij}(\theta) - \Tcosebi{n\-}(\theta) \, \ximT{ij}(\theta)\right].
    \label{eq:COSEBI_from_xipm}
\end{align}
We use the logarithmic COSEBI filters defined by \citet{Asgari2012},  and apply Eq.~\eqref{eq:COSEBI_from_xipm} to the measured correlation functions from simulations and data to obtain the $E_n$ and $B_n$ modes.

For theoretical modelling, it is convenient to express COSEBIs directly in terms of the convergence power spectrum $\CkkT{ij}(\ell)$ for s flat Universe under the Limber approximation and the flat-sky limit,
\begin{equation}
    \CkkT{ij}(\ell) = \int_0^{\chi_H} \frac{\mathrm{d}\chi}{\chi^2} \, q^{(i)}(\chi) \, q^{(j)}(\chi) \, 
    P_\delta\!\left( \frac{\ell+1/2}{\chi}, z(\chi) \right) ,
\end{equation}
where $q^{(i)}(\chi)$ is the lensing efficiency of tomographic bin $i$, $\chi$ is the co-moving distance, and $P_\delta(k,z)$ is the matter power spectrum, which we model using the BACCO emulator \citep{Angulo2021}. The COSEBI modes can then be written as
\begin{equation}
    \EnT{n}{ij} = \int_0^\infty \frac{\mathrm{d}\ell \, \ell}{2\pi} \, \CkkT{ij}(\ell) \, W_{n}^\mathrm{cos}(\ell), 
    \qquad \BnT{n}{ij} = 0,
\end{equation}
where $W_{n}^\mathrm{cos}(\ell)$ are the Fourier-space window functions derived from the real-space filters $\Tcosebi{n\pm}(\theta)$.

\subsection{Third-order aperture statistics}

As our three-point weak lensing statistic we use the third-order aperture mass moments, $\langle \Map ^3 \rangle$, which were employed in the KiDS-1000 analysis of \citet{Burger2024} and whose modelling has been described in detail by \citet{Heydenreich2023}.\footnote{Our full modelling code is available at \url{https://github.com/llinke1/threepoint_py/releases/tag/v1.0.0} \citep{threepoint_py}. } These statistics are defined as the third-order moment of the aperture mass $M_\mathrm{ap}(\boldsymbol{\vartheta};\theta)$, 
\begin{equation}
    \langle \Map ^3\rangle^{(ijk)}(\theta_1,\theta_2,\theta_3) \equiv 
    \big\langle \Map^{(i)}(\boldsymbol{\vartheta};\theta_1)\,
    \Map^{(j)}(\boldsymbol{\vartheta};\theta_2)\,
    \Map^{(k)}(\boldsymbol{\vartheta};\theta_3)\big\rangle ,
\end{equation}
where $\Map^{(i)}(\boldsymbol{\vartheta};\theta_1)$ is the aperture mass for tomographic bin $i$ at position $\boldsymbol{\vartheta}$ smoothed on scale $\theta$. One can consider both equal-scale $\langle \Map ^3 \rangle$ for which $\theta_1 = \theta_2 = \theta_3$, and unequal-scale $\langle \Map ^3 \rangle$ for which the three scales have independent values. 

The aperture mass is defined as a convolution of the convergence field $\kappa$ with a compensated filter function $U$,
\begin{equation}
    \Map^{(i)}(\boldsymbol{\vartheta};\theta) = \int \mathrm{d}^2\vartheta' \,
    U_\theta(|\boldsymbol{\vartheta}'-\boldsymbol{\vartheta}|)\, 
    \kappa^{(i)}(\boldsymbol{\vartheta}') ,
\end{equation}
where $U_\theta$ integrates to zero, ensuring that $\Map$ is insensitive to constant mass sheets.  
Alternatively, $\Map$ can be obtained %directly 
from the tangential shear $\gamt$ using the corresponding shear filter $Q_\theta$,
\begin{equation}
    \Map^{(i)}(\boldsymbol{\vartheta};\theta) = \int \mathrm{d}^2\vartheta' \,
    Q_\theta(|\boldsymbol{\vartheta}'-\boldsymbol{\vartheta}|)\,
    \gamt^{(i)}(\boldsymbol{\vartheta}';\boldsymbol{\vartheta}),
\end{equation}
where $Q_\theta$ is related to $U_\theta$ via
\begin{equation}
    Q_\theta(\vartheta) = \frac{2}{\vartheta^2} \int_0^\vartheta \mathrm{d}\vartheta'\, \vartheta'\, 
    U_\theta(\vartheta') - U_\theta(\vartheta) .
    \label{eq:UQrelation}
\end{equation}

As shown by \citet{Schneider2005}, $\langle \Map^3 \rangle$ can be obtained from the third-order shear correlation functions $\Gamma_n$ , defined as
\begin{align}
    \label{eq:Gamma_definitions}
    &\Gamma_0^{(ijk)} = \langle \gamma^i(\boldsymbol{\vartheta}_1)\, 
                    \gamma^j(\boldsymbol{\vartheta}_2)\,
                    \gamma^k(\boldsymbol{\vartheta}_3)\rangle \,, 
    &\Gamma_1^{(ijk)} = \langle \gamma^{i*}(\boldsymbol{\vartheta}_1)\, 
                    \gamma^j(\boldsymbol{\vartheta}_2)\,
                    \gamma^k(\boldsymbol{\vartheta}_3)\rangle \;, \\
    &\notag \Gamma_2^{(ijk)} = \langle \gamma^{i}(\boldsymbol{\vartheta}_1)\, 
                    \gamma^{j*}(\boldsymbol{\vartheta}_2)\,
                    \gamma^{k}(\boldsymbol{\vartheta}_3)\rangle \,, 
    &\Gamma_3^{(ijk)} = \langle \gamma^{i}(\boldsymbol{\vartheta}_1)\, 
                    \gamma^{j}(\boldsymbol{\vartheta}_2)\,
                    \gamma^{k*}(\boldsymbol{\vartheta}_3)\rangle .
\end{align}

For theoretical modelling, $\langle M_\mathrm{ap}^3\rangle$ can equivalently be written in terms of the projected matter bispectrum $b(\boldsymbol{\ell}_1,\boldsymbol{\ell}_2,\boldsymbol{\ell}_3)$, which in the Limber approximation reads
\begin{align}
     \label{eq:bispectrum_Limber}
    b^{(ijk)}(\boldsymbol{\ell}_1,\boldsymbol{\ell}_2,\boldsymbol{\ell}_3) 
    &= \int_0^{\chi_{\rm H}} \frac{\mathrm{d}\chi}{\chi^4} \, 
     q^{(i)}(\chi) \, q^{(j)}(\chi) \,q^{(k)}(\chi)\\
    & \notag \times B_\delta\left(\frac{\boldsymbol{\ell}_1}{\chi},\frac{\boldsymbol{\ell}_2}{\chi},\frac{\boldsymbol{\ell}_3}{\chi};z(\chi)\right)\;,
\end{align}
with $q^{(i)}(\chi)$ the lensing efficiency of bin $i$ and $B_\delta$ the matter bispectrum.  
The aperture statistic is then given by
\begin{align}
    \langle \Map^3\rangle^{(ijk)}(\theta_1,\theta_2,\theta_3) &=
    \int \frac{\mathrm{d}^2\ell_1}{(2\pi)^2}\frac{\mathrm{d}^2\ell_2}{(2\pi)^2}
    \, b^{(ijk)}(\boldsymbol{\ell}_1,\boldsymbol{\ell}_2,-\boldsymbol{\ell}_1-\boldsymbol{\ell}_2)\, \\
    &\notag \quad \times
    \hat{U}(\ell_1\theta_1)\,\hat{U}(\ell_2\theta_2)\,
    \hat{U}(|\boldsymbol{\ell}_1+\boldsymbol{\ell}_2|\theta_3),
    \label{eq:Map3_from_bispectrum}
\end{align}
where $\hat{U}$ is the Fourier transform of the aperture filter.  By exchanging the order of integration over comoving distance $\chi$ and multipoles $\ell$, we can write
\begin{align}
    &\langle \Map^3\rangle^{(ijk)}(\theta_1,\theta_2,\theta_3) \\ 
    &\notag =
    \int_0^{\chi_H} \frac{\mathrm{d}\chi}{\chi^4} \, 
    q^{(i)}(\chi) q^{(j)}(\chi) q^{(k)}(\chi)\,    \int \frac{\mathrm{d}^2\ell_1}{(2\pi)^2}\frac{\mathrm{d}^2\ell_2}{(2\pi)^2} 
    \\
    &\notag \quad \times
    B_\delta\!\left(\frac{\boldsymbol{\ell}_1}{\chi},\frac{\boldsymbol{\ell}_2}{\chi},\frac{\boldsymbol{\ell}_3}{\chi};z(\chi)\right)
\hat{U}(\ell_1\theta_1)\,\hat{U}(\ell_2\theta_2)\,
    \hat{U}(|\boldsymbol{\ell}_1+\boldsymbol{\ell}_2|\theta_3),\\
    &\notag = \int_0^{\chi_H} \frac{\mathrm{d}\chi}{\chi^4} \, 
    q^{(i)}(\chi) q^{(j)}(\chi) q^{(k)}(\chi)\,\langle M_\mathrm{ap}^3\rangle_z(\theta_1,\theta_2,\theta_3;\chi)\;,
    \label{eq: definition Map3z}
\end{align}
where $\langle \Map^3\rangle_z$ denotes the contribution from %a fixed 
lens redshift $z(\chi)$. 

To accelerate cosmological inference, \citet{Burger2024} trained a neural network emulator for $\langle \Map^3\rangle$ over the KiDS 1000 redshift distribution. However, that emulator implicitly assumed a fixed source distribution and intrinsic alignment model, limiting its applicability to other surveys.  
We instead follow the approach of \citet{Gomes2024} and train an emulator for $\langle \Map^3\rangle_z$. This decouples the redshift distribution from the emulator and allows us to flexibly predict aperture statistics for arbitrary source samples, enabling applications not only to KiDS but also to upcoming surveys such as \Euclid and LSST.  

For the matter bispectrum $B_\delta$ we adopt the \textsc{BiHalofit} model of \citet{Takahashi2020}, which provides an analytic fit to $N$-body simulations across a wide range of scales and redshifts.
We use the exponential compensated filter introduced by \citet{Crittenden2002}, given by
\begin{equation}
    U_\theta(\vartheta) = \frac{1}{2\pi \theta^2} 
    \left( 1 - \frac{\vartheta^2}{2\theta^2} \right) 
    \exp\!\left(-\frac{\vartheta^2}{2\theta^2}\right),
\end{equation}
with 
\begin{align}
    &Q_\theta(\vartheta) = \frac{\vartheta^2}{4\pi \theta^4} 
    \exp\!\left(-\frac{\vartheta^2}{2\theta^2}\right),
    &\hat{U}(\ell\theta) = \frac{(\ell\theta)^2}{2} 
    \exp\!\left(-\frac{(\ell\theta)^2}{2}\right).
\end{align}

% \begin{figure}
%     \centering
%     \includegraphics[width=\linewidth]{Figs/output.png}
%     \caption{Caption}
%     \label{fig:enter-label}
% \end{figure}

\subsection{Shape and redshift calibration biases}

Weak lensing cosmology requires precise calibration of both galaxy shape measurements and photometric redshift distributions. Residual biases in these calibrations can propagate directly into cosmological parameter estimates and must therefore be modelled explicitly.  

Biases in shear measurements are commonly parametrized by a multiplicative term $m$ and an additive term $c$, such that
\begin{equation}
    \gamma_\mathrm{obs} = (1+m)\,\gamma_\mathrm{true} + c ,
\end{equation}
where $\gamma_\mathrm{obs}$ denotes the measured shear and $\gamma_\mathrm{true}$ the underlying cosmic shear signal. The determination of $m$ and correction of $c$ for KiDS-Legacy is described in \citet{Wright2025}. In our analysis we marginalise over the multiplicative bias by including $m$ as a free nuisance parameter in the modelling. Its value is drawn from an informative prior calibrated from image simulations \citep{Li2023}. The additive bias $c$ is corrected for in the catalogue construction.  

Photometric redshift uncertainties can also bias the inferred shear signal. A leading effect is a systematic shift in the mean of the redshift distributions $n^{(i)}(z)$ for each tomographic bin $i$. We model this as
\begin{equation}
    n_\mathrm{true}^{(i)}(z) = n_\mathrm{obs}^{(i)}(z+\delta z^{(i)}),
\end{equation}
where $\delta z^{(i)}$ is a shift parameter for each tomographic bin. These $\delta z^{(i)}$ parameters are included in the inference and constrained with Gaussian priors derived from the dedicated redshift calibration of KiDS-Legacy \citep{Wright2025a}.  

\subsection{Intrinsic alignments}
\label{sc:Model:IA}
Weak lensing analyses must account for the IA of galaxy shapes, which contaminate the cosmic shear signal. In general, the tomographic cosmic shear power spectrum, including IA, can be written as
\begin{equation}
    C^{(ij)}(\ell) = C_\mathrm{GG}^{(ij)}(\ell) + C_\mathrm{GI}^{(ij)}(\ell) + + C_\mathrm{GI}^{(ji)}(\ell) + C_\mathrm{II}^{(ij)}(\ell),
\end{equation}
where $C_\mathrm{GG}$ is the pure lensing signal, $C_\mathrm{GI}$ the correlation between gravitational shear and intrinsic shapes, and $C_\mathrm{II}$ the intrinsic shape auto-correlation.  
The IA terms can be expressed in the Limber approximation as
\begin{align}
    C_\mathrm{GI}^{(ij)}(\ell) &= \int_0^{\chi_H} \frac{\mathrm{d}\chi}{\chi^2}\,
    q^{(i)}(\chi)\, n^{(j)}(\chi)\, P_\mathrm{GI}\!\left(\frac{\ell}{\chi},z(\chi)\right), \\
    C_\mathrm{II}^{(ij)}(\ell) &= \int_0^{\chi_H} \frac{\mathrm{d}\chi}{\chi^2}\,
    n^{(i)}(\chi)\, n^{(j)}(\chi)\, P_\mathrm{II}\!\left(\frac{\ell}{\chi},z(\chi)\right),
    \label{eq:Limber_IA}
\end{align}
where $n^{(i)}(\chi)$ is the source comoving distance distribution in bin $i$ and $P_\mathrm{GI}, P_\mathrm{II}$ are the intrinsic alignment 3D power spectra \citep{Hirata2004,Bridle2007}.   

Extending this formalism to three-point statistics yields four bispectrum contributions,
\begin{equation}
    b^{(ijk)} = b_\mathrm{GGG}^{(ijk)} + b_\mathrm{GGI}^{(ijk)} + b_\mathrm{GII}^{(ijk)} + b_\mathrm{III}^{(ijk)} + \textrm{perms of $i,j,k$} \;,
\end{equation}
where the terms denote the pure lensing bispectrum and the various IA-contaminated contributions. For example,
\begin{align}
    b_\mathrm{GGI}^{(ijk)}(\boldsymbol{\ell}_1,\boldsymbol{\ell}_2,\boldsymbol{\ell}_3) &=
    \int_0^{\chi_H} \frac{\mathrm{d}\chi}{\chi^4}\,
    q^{(i)}(\chi)\, q^{(j)}(\chi)\, n^{(k)}(\chi)\, \\
    &\notag \quad \times
    B_\mathrm{GGI}\!\left(\frac{\boldsymbol{\ell}_1}{\chi},\frac{\boldsymbol{\ell}_2}{\chi},\frac{\boldsymbol{\ell}_3}{\chi};z(\chi)\right),
\end{align}
and analogously for $b_\mathrm{GII}$ and $b_\mathrm{III}$.  
Different IA models specify the forms of $P_\mathrm{GI}$, $P_\mathrm{II}$, and their bispectrum counterparts $B_\mathrm{GGI}, B_\mathrm{GII}, B_\mathrm{III}$.

Following \citet{Burger2024}, we adopt a Non-linear Linear Alignment (NLA)-inspired description for the bispectrum IA terms, with
\begin{equation}
    B_\mathrm{GGI} = f_\mathrm{IA}\, B_\mathrm{GGG}, \quad
    B_\mathrm{GII} = f_\mathrm{IA}^2\, B_\mathrm{GGG}, \quad
    B_\mathrm{III} = f_\mathrm{IA}^3\, B_\mathrm{GGG}.
\end{equation}
\citet{Linke2024} demonstrated that such a scaling provides a good fit to the observed third-order intrinsic correlations of massive, low-redshift galaxies.  We note, however, that recently \citet{Vedder2026} found a significantly worse agreement of NLA models to simulated galaxies in the FLAMINGO simulations.
Unlike \citet{Burger2024} we do not assume a purely scale-independent constant prefactor $f_\mathrm{IA}$. Instead, we follow the KiDS-Legacy second-order analysis of \citet{Wright2025} and implement three variants of the NLA model,

\begin{equation}
        f_\mathrm{IA}(z) = - A_{\rm IA}\;
        \frac{C_1 \rho_{\rm cr}\, \Omega_{\rm m}}{D(z)} , \hspace{0.5cm} \mathrm{(NLA)}
\end{equation}

\begin{equation}
        f_\mathrm{IA}(z) = - \left[ A_{\rm IA} + b_{\rm IA} 
        \left( \frac{\langle a \rangle^{(i)}}{a_{\rm piv}} -1 \right) \right]
        \frac{C_1 \rho_{\rm cr}\, \Omega_{\rm m}}{D(z)} , \hspace{0.5cm} \mathrm{(NLA-}z\mathrm{)}
\end{equation}
and
\begin{equation}
        f_\mathrm{IA}(z,M_h) = - A_{\rm IA}\; f_{\mathrm{r}}^{(i)}\;
        \left( \frac{\langle M_{\rm h} \rangle^{(i)}}{M_{\rm h,piv}} \right)^\beta
        \frac{C_1 \rho_{\rm cr}\, \Omega_{\rm m}}{D(z)} . \hspace{0.5cm} \mathrm{(NLA-}M\mathrm{)}
\end{equation}

Here $C_1$ is a normalisation constant \citep{Bridle2007}, $\rho_\mathrm{cr}$ is the critical matter density, $D(z)$ is the linear growth factor, $\langle a \rangle^{(i)}$ the mean scale factor of sources in tomographic bin $i$, $a_\mathrm{piv}$ a pivot scale factor, $\langle M_{\rm h}\rangle^{(i)}$ the average halo mass of sources, and $f_\mathrm{r}^{(i)}$ the red galaxy fraction.  

In contrast to \citet{Wright2025}, we do not employ the "NLA-$k$" model, which incorporates a scale-dependent IA amplitude. The extension of this model to higher-order statistics is not straightforward: while the power spectrum depends on a single scale $k$, allowing for a natural parametrization of scale dependence, the bispectrum depends on three scales ($k_1$, $k_2$, $k_3$). How the scale dependence should generalise to this configuration space remains ambiguous.

Our fiducial choice is the NLA-$M$ model, which allows for a redshift- and mass-dependent IA amplitude tied to galaxy populations, in line with the 2-point analysis of \citet{Wright2025}. This parametrization captures the observed dependence of intrinsic alignments on galaxy type, halo mass, and redshift evolution, while remaining tractable for cosmological inference. We adopt this model as fiducial as it is the most flexible one of our choices, and naturally accounts for the fact that different tomographic bins have different galaxy type compositions, which is neglected by the NLA and NLA-$z$ models. 

\subsection{Baryonic feedback}
Accurate modelling of baryonic feedback is essential to exploit small-scale information in cosmic shear and higher-order statistics. Baryonic processes such as gas cooling, star formation, and AGN feedback redistribute matter within haloes, leading to a suppression of the matter power spectrum and bispectrum on non-linear scales \citep{Semboloni2013}.  

While \citet{Burger2024} modelled baryonic effects using suppression functions calibrated on the \textsc{Magneticum} simulations, we now employ a consistent baryon correction model for both power spectrum and bispectrum developed by \citet{Burger2025}. This model introduces 5 free parameters that characterise how baryonic physics modifies halo properties such as concentration, gas fraction, and stellar content. It was trained on dark-matter-only simulations that were baryonified using the BACCO approach \citep{Angulo2021}, and subsequently validated against the hydrodynamical FLAMINGO simulations \citep{Schaye2023, Kugel2023}. \citet{Burger2025} showed that for cosmic shear, the most important quantities are the mass of hot gas, parametrized by $M_\mathrm{c}$ and the density of ejected gas, parametrized by $\eta$. We therefore restrict our analysis to varying only these two parameters, holding the remaining three fixed at their fiducial values. While this reduces model flexibility compared to the full five-parameter space, it represents a pragmatic choice given the computational cost of sampling and the finding that these two parameters capture the dominant baryonic effects on cosmic shear.

Our baryonic suppression model is no longer tied to a single hydrodynamical simulation but instead interpolates between multiple feedback scenarios, ensuring a physically motivated and flexible description of baryonic effects for both the power spectrum and bispectrum.

Other analyses (e.g. \citealp{Sugiyama2025}, \citealp{Gomes2025}) have opted to not model the impact of baryonic effects and instead restricted their analysis to scales where the impact of baryons is small. However, for our data set this would require removing a large part of our data vector. As we show in Appendix \ref{app: impact baryons}, realistic baryonic impact (taken from the FLAMINGO simulation suite) can impact the data vector even for aperture radii as large as \ang{;10;} by more than 30\% of the statistical error. As third-order statistics contribute the most information at scales below this value, it is vital for us to include a baryonification model.

Additionally, recent works \citep[e.g.,][]{2024MNRAS.534..655B,2025PhRvD.112h3509H} provide evidence that baryon feedback is stronger and acts on larger scales than previously assumed, hinting that scale-cuts derived by state-of-the-art hydrodynamical simulations may be overly optimistic. This tentative conclusion is also supported by \citet{Arico2023}, underscoring the need for a flexible model of baryonic processes.

\subsection{Reduced shear and source clustering}

Cosmic shear modelling is affected by higher-order effects that arise when moving beyond the weak lensing approximation. In principle, these corrections introduce terms involving the bispectrum, trispectrum, and higher-order correlators into the expression for the shear power spectrum $C(\ell)$ (see, e.g., \citealt{Deshpande-EP28} for an overview). Among these effects, the most relevant for our analysis are the reduced shear approximation and source clustering, both of which can bias cosmological inference if neglected.  

The observed galaxy ellipticity $\epsilon$ can be written as the sum of the intrinsic ellipticity $\epsilon_S$ and the reduced shear $g$, 
\begin{equation}
    \epsilon \simeq \epsilon_S + g, \qquad 
    g = \frac{\gamma}{1-\kappa},
\end{equation}
where $\gamma$ is the shear and $\kappa$ the convergence. In most analyses one applies the approximation $g \simeq \gamma$, which neglects terms of order $\gamma\kappa$ and higher. This reduced shear approximation leads to a systematic bias that increases towards smaller angular scales and higher redshift.  

A second effect arises from the fact that estimators for shear correlation functions typically assume that source galaxies are uniformly distributed and uncorrelated with the underlying matter field. In reality, galaxies cluster in over-dense regions, which are also regions of stronger gravitational lensing. This induces correlations between source positions and shear, known as {source clustering}. While negligible for second-order cosmic shear statistics \citep{Linke2025}, the effect can become significant for higher-order statistics \citep{Gatti2024}, and thus must be accounted for when modelling the bispectrum and aperture mass statistics.  

We model both reduced shear and source clustering using simulated shear and convergence maps (see Sect.~\ref{sc:Data:Simulation}). From these simulations we construct data vectors for $E_n$ and $\langle M_\mathrm{ap}^3\rangle$ both including and excluding the higher-order effects. The ratio of the two then provides a multiplicative correction factor for each data point. In our fiducial analysis, the measurements are corrected by dividing by this ratio, thereby removing the combined impact of reduced shear and source clustering.  

\begin{figure*}
    \centering
    \includegraphics[width=\linewidth]{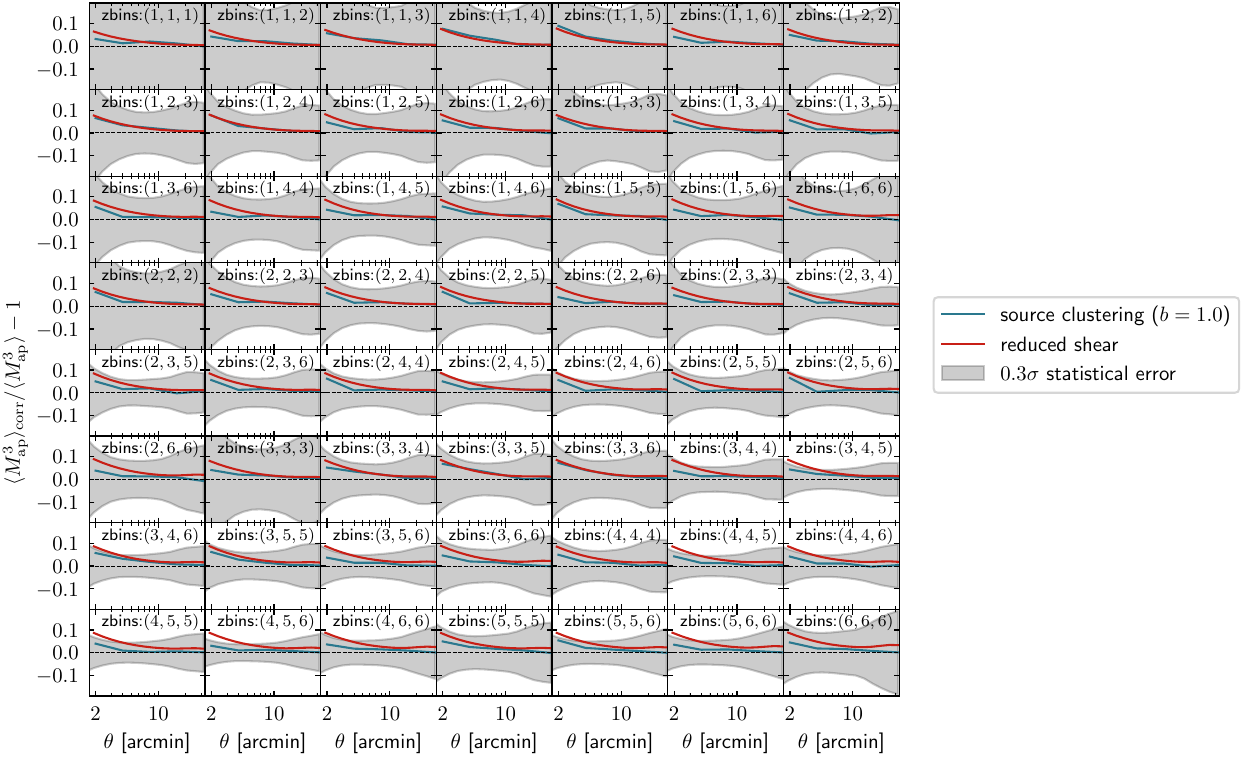}
    \caption{Relative error of $\langle M_\mathrm{ap}^3\rangle_\mathrm{corr}(\theta, \theta, \theta)$ with reduced shear (red) or source clustering (blue) to the $\langle M_\mathrm{ap}^3\rangle(\theta, \theta, \theta)$ without these effects. The grey-filled area is $0.3$ of the statistical uncertainty from the covariance matrix estimate. Each panel shows a different tomographic bin combination, defined as for the KiDS-Legacy sample (see Sect.\ref{sc:Data}) }
    \label{fig: impact higher order}
\end{figure*}

Figure~\ref{fig: impact higher order} illustrates the impact of reduced shear and source clustering on $\langle M_\mathrm{ap}^3\rangle$ for a fixed galaxy bias $b=1$ in the equal-scale  case. Both effects show a stronger impact at small scales, with the reduced shear being the dominant effect. However, the impact of source clustering is of the same order of magnitude, so either none or both of the effects need to be taken into account in an analysis. The impacts consistently lie below 30\% of the statistical uncertainty of the measurements. %However, since the effects always contribute positively, they can lead to biases in the inferred cosmology and we consequently include them in our modelling.

\section{\label{sc:Data} Data}
\subsection{Takahashi simulations}
\label{sc:Data:Simulation}
We use simulated data for two purposes: modelling the covariance of our data, and estimating the impact of reduced shear and source clustering.
Our simulated data are based on the full-sky gravitational lensing simulations by \citet{Takahashi2017} (T17 in the following), used also in \citet{Burger2024}.
These N-body simulations evolve $2048^3$ particles in a series of nested cosmological volumes and thus produce 108 independent full-sky realisations of the shear and convergence maps for 38 different redshift slices.
The simulations use the cosmological parameters
$\Omm=0.279$, $\OmLa=1-\Omm$, $h=0.7$, $n_\mathrm{s}=0.97$, $\Omb=0.046$, $\sigma_8=0.82$.

We create three types of data sets from these simulations. 
The first type is used to estimate the covariance. For this, we cut out 18 footprints of the shape of KiDS-Legacy from each full-sky map. We then create simulated catalogues from these maps by taking their shear value at the positions of observed KiDS legacy galaxies. To match the shape noise of the true data, we add the observed ellipticities (randomly rotated) to the simulated shear. 

The other two data sets are used to assess (and correct for) the impact of the reduced shear approximation and source clustering. For both we use a single full sky map without including any shape noise, to increase our sensitivity to these subtle effects. The data set without source clustering is created by distributing source galaxies uniformly across each redshift slice, such that the galaxy number density follows the KiDS-Legacy $n(z)$. To include the impact of source clustering, we use the density contrast $\delta$ at each redshift slice of the simulation. For each pixel $\Vec{x}$ we draw $N [1+b \delta(\Vec{x})]$ galaxies, where $b$ is an assumed galaxy bias and $N$ chosen such that the mean galaxy number follows the KIDS-Legacy $n(z)$. One can easily imagine that the impact of source clustering depends on the choice of the galaxy bias $b$. However, determining this bias for a weak lensing source sample, which is subject to many subtle selection effects, is not straightforward. Nevertheless, as we show later, the impact of source clustering is small for our statistics, and thus the concrete choice is not critical for our analysis. We opt for a fiducial value of $b=1$ (as was also assumed e.g. in \citealt{Gatti2024}) and two alternative values $b=0.7$ and $b=1.3$.

Assessing the impact of the reduced shear approximation is comparatively easy. We use the simulated catalogue without source clustering and measure the cosmic shear statistics for either the shear $\gamma$ or the reduced shear $g=\gamma / (1+\kappa)$. The ratio between both measurements gives the necessary correction.

\subsection{KiDS-Legacy}

Our %cosmological 
analysis is based on the KiDS-Legacy weak lensing sample, described in detail by \citet{Wright2025}. 
It is drawn from the fifth and final data release (DR5) of KiDS \citet{Wright2024}, which provides 1347 deg$^2$ of imaging in the $u$, $g$, $r$, and $i$ bands obtained with OmegaCAM \citep{Kuijken2011} at the VLT Survey Telescope between 2011 and 2019, complemented by near-infrared $ZYJHK_s$ observations from the VIKING survey \citep{Edge2013}. 

\begin{figure}
    \centering
    \includegraphics[width=\linewidth]{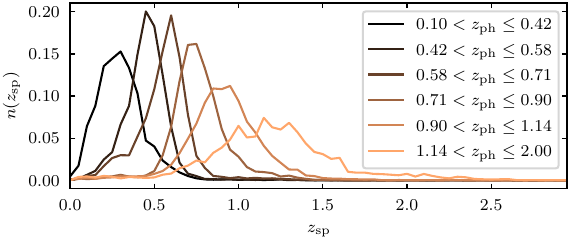}
    \caption{Redshift distribution of KiDS-Legacy galaxies per tomographic bin.}
    \label{fig:nz}
\end{figure}

Compared to the previous release (DR4; \citealt{Kuijken2019}), DR5 covers 34\% more area and increases the $i$-band depth through a second pass. The number of spectroscopic sources available for photometric-redshift calibration has increased by a factor of five, extending the photometric redshift range of the lensing sample from $z<1.2$ to $z<2$ and increasing the effective survey volume by a factor of 3.5. 
These improvements enabled the definition of six tomographic bins for cosmic shear (instead of five as for KiDS-1000), whose calibrated redshift distributions are shown in Fig.~\ref{fig:nz}. The  tomographic bins are defined by the folloing intervals in photometric redshift $z_\mathrm{B}$: $ (0.10, 0.42], (0.42, 0.58], (0.58, 0.71], (0.71, 0.90], (0.90, 1.14]$ and $(1.14, 2.00]$.

Weak lensing shapes are measured with an updated version of the \textsc{lensfit} algorithm \citep{Miller2013}, including improved sampling of the ellipticity likelihood and corrections for PSF leakage \citep{Li2023}. Additional updates to photometric calibration and masking are described in \citet{Wright2025}. 
Originally, a residual astrometric issue was identified for the KiDS-Legacy data set in $B$-mode tests. Subsequently, approximately 4\% of sources impacted strongest by this issue were excluded. 
The resulting effective lensing sample spans $967.4 \,\mathrm{deg}^2$ after masking, with an effective galaxy number density of $n_\mathrm{eff} = 8.79 \,\mathrm{arcmin}^{-2}$. Aside from the shapes, \textsc{lensfit} also provides a weight for each source galaxy, with galaxies with more reliable shape measurements receiving a higher weight. This weight is automatically taken into account in the correlation function measurement.

\section{\label{sc:Methods} Methods}
\subsection{Measurement}

We obtain $E_n$ and $\langle \Map^3\rangle$ from the observed and simulated data from the second- and third-order shear correlation functions $\xipm(\vartheta)$ and $\Gamma_n(\vartheta_1, \vartheta_2, \phi)$, where $\vartheta$ is the angular separation between two galaxies in a pair, $\vartheta_1$ and $\vartheta_2$ are the angular side lengths of a triangle spanned by a triplet of galaxies, and $\phi$ the opening angle of this triangle. For the simulations, we measure the $\xi_\pm$ using \verb|treecorr| \citep{Jarvis2004} with the setup from \citet{Wright2025}. For the observed data, we use the $E_n$ measurements from \citet{Wright2025} directly. 

To estimate the third-order correlation functions, we use the code \verb|orpheus| \citep{Porth2024, Porth2025}\footnote{\url{https://github.com/lporth93/orpheus/}}, that employs a multipole decomposition of the correlation functions for an optimised evaluation. The third-order correlation functions are measured for  $\vartheta_1, \vartheta_2 \in [\ang{;0.5;}, \ang{;256;}]$ with logarithmic bin width $0.15$. We evaluate the multipoles up to a maximum multipole of $N_\mathrm{max}=30$ and use 128 linearly spaced bins for $\phi$ for the transformation of the correlation functions to the aperture statistics $\langle M_\mathrm{ap}^3 \rangle$. To speed up the calculation we used the DoubleTree approximation (see Appendix F3 in \citealt{Porth2025}) using a hierarchy of reduced catalogues of pixel sizes $[\ang{;0.5;}, \ang{;1;}, \ang{;2;}]$ and set the \verb|rmin_pixsize| parameter to 15. Using this setup we measure all 216 tomographic bin combinations of the third-order correlation functions in 250 CPUh on an AMD EPYC 9554 processor.

\subsection{Blinding}

To prevent confirmation bias in the interpretation of our results, we implemented a strict blinding procedure for the KiDS-Legacy third-order shear analysis. A designated “unblinder” outside of the core analysis team was responsible for creating a PGP key pair and providing only the public key to the team. The unblinded data vector was read exactly once and immediately encrypted with this key, ensuring that no team member had access to the true data during the analysis.  

The blinding proceeded as follows. At a chosen reference cosmology $C$ we computed $\langle M_\mathrm{ap}^3\rangle(C)$. A set of three random shifts $\Delta_i$ in the cosmological parameters $\sigma_8$ and $\Omega_\mathrm{m}$ was generated, with the list of shifts stored in a file encrypted by the unblinder’s public key. For each shift, we calculated the difference $\langle M_\mathrm{ap}^3\rangle(C) - \langle M_\mathrm{ap}^3\rangle(C+\Delta_i)$ and subtracted it from the unblinded data vector to obtain a blinded version. The blinded data vectors were provided to the analysis team and used for all methodological development, validation tests, and parameter inference.  

The set of shifts included the trivial case $\Delta_i = 0$, so that the true data vector was analysed alongside the blinded variants. However, identifying which case corresponded to the real data required decrypting either the parameter-shift file or the encrypted unblinded data vector, a step that could only be performed by the unblinder with the private key. This ensured that the analysis remained fully blind until all methodological decisions were finalised and the unblinder authorised decryption.

\subsection{Emulators}
Sampling the posterior distribution requires thousands of model evaluations. While this is routinely feasible for second-order statistics, the evaluation of the third-order aperture mass $\langle M_\mathrm{ap}^3 \rangle$ involves the triple integral of Eq.~\eqref{eq:Map3_from_bispectrum}, which makes a direct computation prohibitively time-consuming. Even with optimised numerical integration, evaluating $\langle M_\mathrm{ap}^3 \rangle$ takes several minutes per parameter set, which is incompatible with Markov Chain Monte Carlo or Hamiltonian sampling.  

To overcome this challenge, we develop a neural network emulator for the redshift-resolved statistic $\langle M_\mathrm{ap}^3 \rangle_z$, with only the Limber projection performed in each model evaluation. This reduces the evaluation time from minutes to milliseconds on a single CPU core, enabling efficient cosmological inference. In practice, we further parallelise the inference over tens of CPU cores.

Our emulator is implemented using the \textsc{cosmo-emu-jax} framework\footnote{\url{https://github.com/pburger112/cosmoemuJAX}}, which provides tools for creating, training, and validating neural network emulators. The framework supports flexible architectures with arbitrary input and output dimensions, multiple fully-connected hidden layers, a range of activation and loss functions, and customisable training hyper-parameters. 

The training set consists of 20\,000 evaluations of $\langle \Map^3 \rangle_z$ sampled in a Latin hypercube design across the parameters $S_8$, $\Omega_\mathrm{m}$, $\Omega_\mathrm{b}$, $n_S$, $h$, $M_\mathrm{c}$, and $\eta$. The emulator is trained on a randomly selected subset of 85 \% of these evaluations, while accuracy is validated on the remaining samples. To optimise the network architecture, we use Bayesian optimisation of the hyperparameters with the \texttt{optuna} package \citep{Akiba2019}. The resulting optimal configuration (see Appendix \ref{app: emulators}) yields an accuracy of better than 2\% (4\%) for the 95\% (99\%) percentile (see Fig.~\ref{fig: Map3 emu accuracy}).  

In addition to $\langle M_\mathrm{ap}^3 \rangle_z$, we train emulators for the COSEBI modes $E_n$. Ideally, one would emulate the functions $E_{n,z}(\chi)$, which are independent of the source redshift distribution and included in
\begin{equation}
    \EnT{n}{ij} = \int_0^{\chi_H} \frac{\mathrm{d}\chi}{\chi^2} \, q^{(i)}(\chi) \, q^{(j)}(\chi) \, E_{n,z}(\chi)\;.
\end{equation}
However, $E_{n,z}$ contains oscillatory structures that are smoothed out in the Limber projection but must be captured with high precision to recover accurate $E_n$. We found these oscillations prohibitively difficult for a neural network to learn even with deep architectures, leading to unstable predictions.  

Consequently, we train three separate emulators directly for $E_n$, each tailored to the KiDS-Legacy redshift distribution and assuming a different IA model (NLA, NLA-$z$, and NLA-$M$; see Sect.~\ref{sc:Model:IA}). This approach ensures accurate predictions for all IA scenarios considered in our analysis. Shifts in the mean of the $n(z)$ are incorporated in the emulator by extending the latin hypercube of training parameters to include varying $\delta z$. The performance of the COSEBI emulators is illustrated in Fig.~\ref{fig: En emu accuracy}, where the residuals with respect to the exact calculation for the NLA-$M$ model are shown to remain within a few per cent. Appendix \ref{app: emulators} lists the optimal network architectures for the COSEBI emulators.

\subsection{Covariance construction}
\label{sc:methods:covariance}

Reliable parameter inference requires an accurate description of the covariance matrix of the data vector. We estimate the covariance from mock survey realisations generated from the T17 simulations. These mocks do not include galaxy clustering, so that the resulting covariance captures the contributions from cosmic variance, shape noise, and survey geometry, but remains free from correlations induced by intrinsic clustering, as well as intrinsic alignments and baryonic feedback. The sample covariance is computed from $N_\mathrm{sim}=1944$ realisations as
\begin{equation}
    \mathbf{C}_\mathrm{sample} = 
    \frac{1}{N_\mathrm{sim}-1} 
    \sum_{i=1}^{N_\mathrm{sim}}
    (\vec{d}_i - \bar{\vec{d}})
    (\vec{d}_i - \bar{\vec{d}})^\mathrm{T}\,,
\end{equation}
where $\vec{d}_i$ denotes the data vector of realisation $i$ and $\bar{\vec{d}}$ the ensemble mean.  

To obtain a covariance matrix that is both accurate and robust for likelihood analyses, the raw sample covariance is rescaled in several steps. The need for rescaling arises because the simulations used to estimate the covariance were run at a fixed cosmology that may differ from the true cosmology of the measurements, meaning the overall amplitude of the sample covariance may not accurately reflect the variance of the data. 

The rescaling procedure proceeds as follows. First, analytic covariance matrices $\mathbf{C}_\mathrm{analytic}$ and $\mathbf{C}_\mathrm{analytic}^{\mathrm{no\,SN}}$ are computed for a reference cosmology, with and without shape noise respectively, following \citet{Linke2023}. The shape noise contribution is then removed from the sample covariance:
\begin{equation}
    \mathbf{C}_\mathrm{no\,SN} = \mathbf{C}_\mathrm{sample} - 
    (\mathbf{C}_\mathrm{analytic} - \mathbf{C}_\mathrm{analytic}^{\mathrm{no\,SN}}).
\end{equation}
Next, the cosmology-dependent part is rescaled. We compute a rescaling matrix from the outer products of the measured data vector $\mathbf{d}$ and the model prediction $\mathbf{m}$ at the simulation cosmology:
\begin{equation}
    \mathbf{R} = \frac{\mathbf{d} \otimes \mathbf{d}}{\mathbf{m} \otimes \mathbf{m}},
\end{equation}
and apply it element-wise to the shape-noise-subtracted covariance:
\begin{equation}
    \mathbf{C}_\mathrm{rescaled} = \mathbf{C}_\mathrm{no\,SN} \odot \mathbf{R},
\end{equation}
where $\odot$ denotes element-wise multiplication and $\otimes$ denotes the outer product. This step anchors the amplitude of the covariance to the observed signal while preserving the correlation structure. Finally, the shape noise is reintroduced analytically:
\begin{equation}
    \mathbf{C}_\mathrm{final} = \mathbf{C}_\mathrm{rescaled} + 
    (\mathbf{C}_\mathrm{analytic} - \mathbf{C}_\mathrm{analytic}^{\mathrm{no\,SN}}).
\end{equation}
This procedure ensures that the cosmology-dependent part of the covariance is properly normalized to the data, while the shape noise contribution remains accurate and unaffected by the rescaling.

The resulting covariance is subsequently inverted for use in the likelihood evaluation. The accuracy of parameter constraints obtained with a finite sample covariance depends on the ratio $N_\mathrm{sim}/N_\mathrm{d}$. While our choice of likelihood, based on \citet{Percival2014}, accounts for noise in the covariance estimate, the formulation requires $N_\mathrm{sim} > N_\mathrm{d} + 4$ to be well-defined. Following the recommendations of \citet{Hartlap2007}, reliable results further require $N_\mathrm{sim} \gg N_\mathrm{d}$ to ensure the inverse of the covariance matrix is well-conditioned. We therefore restrict the data vector to $N_\mathrm{d}\leq 291$ entries, corresponding to 15\% of the available realisations. This choice ensures that both the Percival correction factor $B$ and the effective degrees of freedom $f$ are robustly determined, and that the impact of covariance noise on parameter inference remains small. We discuss the composition of the data vector in Sect. \ref{sc:Results}.

\subsection{Parameter inference}

We derive cosmological parameter constraints within a Bayesian framework. The posterior distribution is given by Bayes' theorem,
\begin{equation}
p(\boldsymbol{\theta}|\vec{d}) \propto
\mathcal{L}(\vec{d}|\boldsymbol{\theta}) \, \pi(\boldsymbol{\theta}),
\end{equation}
where $\boldsymbol{\theta}$ are the model parameters, $\vec{d}$ the data vector, $\mathcal{L}$ the likelihood, and $\pi$ the prior.

The second-order cosmic shear analysis of KiDS Legacy used a Gaussian likelihood for the cosmological parameter inference. However, in contrast to their analysis, our covariance matrix is not calculated analytically but estimated from a finite number of simulations, as described in Sect.~\ref{sc:methods:covariance}. Thus, the standard Gaussian likelihood is biased \citep{Sellentin2016}. To account for this bias, we use instead the likelihood formulation of \citet{Percival2014},
\begin{equation}
\mathcal{L}(\vec{d}|\boldsymbol{\theta}) \propto
\left[1 + \frac{\chi^2(\boldsymbol{\theta})}{N_\mathrm{sim}-1}\right]^{-f/2},
\end{equation}
where
\begin{equation}
\chi^2(\boldsymbol{\theta}) =
(\vec{d}-\vec{m}(\boldsymbol{\theta}))^\mathrm{T}
\mathbf{C}^{-1}
(\vec{d}-\vec{m}(\boldsymbol{\theta})),
\end{equation}
where $\vec{m}(\boldsymbol{\theta})$ is the model prediction and $\mathbf{C}$ the sample covariance matrix estimated from $N_\mathrm{sim}$ realisations. The effective number of degrees of freedom $f$ accounts for the finite number $N_\mathrm{sim}$ of simulations and the number of inferred parameters $N_\mathrm{param}$ as
\begin{equation}
f = N_\mathrm{param} + 2 +
\frac{N_\mathrm{sim} - 1 + B(N_\mathrm{d} - N_\mathrm{param})}
{1 + B(N_\mathrm{d} - N_\mathrm{param})},
\end{equation}
where 
\begin{equation}
B = \frac{N_\mathrm{sim} - N_\mathrm{d} - 2}
{(N_\mathrm{sim} - N_\mathrm{d} - 1)(N_\mathrm{sim} - N_\mathrm{d} - 4)}.
\end{equation}
This approach marginalises over the uncertainty in the covariance estimate and reduces to the Gaussian likelihood in the limit $N_\mathrm{sim}\to\infty$.

We sample over all IA parameters per model, the redshift distribution shifts $\delta z^{(i)}$, the multiplicative shear bias $m^{(i)}$ per tomographic bin $i$, the cosmological parameters $h$, $S_8$, $n_\mathrm{s}$, $\Omm$, $\Omb$, and the baryonic suppression parameters $\eta$ and $M_c$. These two parameters have been found by \citet{Burger2025} to capture the majority of the baryonic suppression.
Our priors are given in Table~\ref{tab: priors}. For the IA, redshift calibration and multiplicative bias parameters,  as well as $h$, $S_8$ and $n_\mathrm{s}$, we adopt the same prior distributions as the KiDS-Legacy second-order cosmic shear analysis. The priors on $\Omega_\mathrm{b}$ and $\Omega_\mathrm{m}$ are also inspired from there, although they assume flat priors in $\Omega_\mathrm{b}h^2$ and $\Omega_\mathrm{m} h^2$. We choose here priors in $\Omega_\mathrm{b}$ and $\Omega_\mathrm{m}$, as both $\langle\Map ^3\rangle$ and the $E_n$ are more sensitive to these parameters and thus the emulators were trained in these parameters. The priors for $\eta$ and $M_c$ are given by the validation range of the baryonic feedback model.

\begin{table}[t]
\centering
\caption{Used prior range}
\begin{tabular}{cc}
\toprule
{Parameter} & {Prior Range} \\
\midrule
$S_8$ & [0.5, 1.0] \\
$\Omega_\mathrm{m}$ & [0.1, 0.68] \\
$\Omega_\mathrm{b}$ & [0.04, 0.06] \\
$n_s$ & [0.84, 1.10] \\
$h$ & [0.64, 0.82] \\
$\log(M_c h / M_\odot)$ & $[10.01, 15.99]^*$ \\
$\eta$ & $[-0.69, 0.19]^*$ \\
$A_1^{\rm IA}, b_1^{\rm IA}$ & $N(\mu_{A_1 b_1}, C_{A_1 b_1})$ \\
$\delta z$ & $N(\mu_{\delta z}, C_{\delta z})^*$\\
$m$ & $N(\mu_{m}, C_{m})^*$\\
\bottomrule
\end{tabular}
\vspace{0.5em}

\small
\textbf{Notes.} Ranges $[$lower, upper$]$ denote the ranges of flat priors. $N(\mu, C)$ denotes Gaussian priors of mean $\mu$ and covariance $C$. The values for the mean and covariances of the redshift calibration and shear bias priors are given in \citet{Wright2025}. We mark with * informative priors.
\label{tab: priors}
\end{table}

Posterior sampling efficiency depends strongly on the choice of algorithm. As our model predictions are provided by neural network emulators implemented in the differentiable JAX framework, we can evaluate not only the likelihood but also its gradients with respect to the parameters. This enables the use of Hamiltonian Monte Carlo (HMC; \citealt{Neal2011}), which explores the posterior more efficiently than random-walk Metropolis–Hastings. Specifically, we employ the No-U-Turn Sampler (NUTS) as implemented in the \texttt{blackjax} library \citep{cabezas2024blackjax}.
%With this sampler, a single MCMC chain of $N$ steps requires XY CPU time for the COSEBI likelihood and XY for the $\langle M_\mathrm{ap}^3 \rangle$ likelihood.  

To ensure convergence, each inference run consists of eight independent chains of length $N$. Convergence is assessed using the Gelman–Rubin statistic \citep{Gelman1992},
\begin{equation}
    \hat{R} = \sqrt{\frac{\widehat{\mathrm{Var}}(\theta|d)}
                        {W}},
\end{equation}
for each inferred parameter $\theta$, where $\widehat{\mathrm{Var}}(\theta|d)$ is the pooled variance across chains and $W$ the within-chain variance. Values of $\hat{R}<1.01$ are taken as evidence for convergence. If this criterion is not met for any parameter, we repeat the analysis with chains of doubled length until convergence is achieved. In practice we find that $N=25\,000$ is enough to ensure convergence for all sampled posteriors.

\section{\label{sc:Results} Results}
\subsection{Measurements of third-order aperture statistics in KiDS-Legacy}
\begin{figure*}
    \centering
    \includegraphics[width=\linewidth]{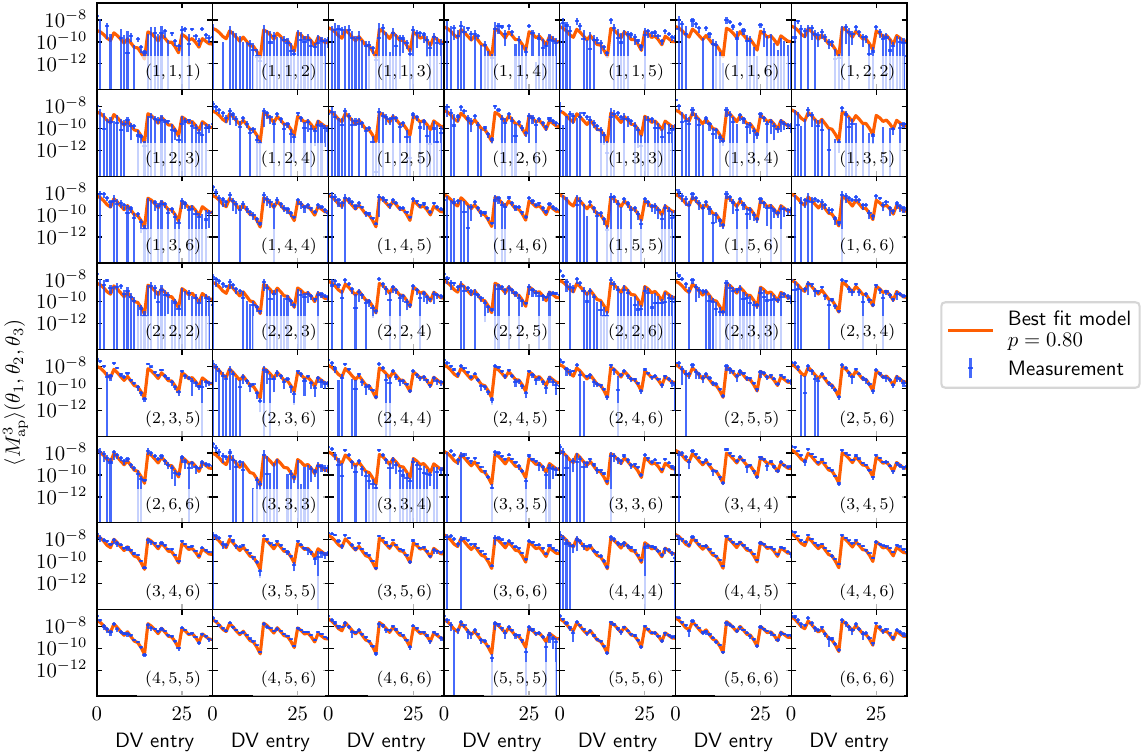}
    \caption{Measured third-order statistics together with best-fit model and $p$-value of best-fit model. The error bars are obtained from the simulation-based covariance estimate.}
    \label{fig: measurements}
\end{figure*}

We show in Fig.\ref{fig: measurements} the third-order aperture statistics measured in KiDS-Legacy together with the best-fitting theoretical model, determined in the fiducial cosmological inference run (see Sect.~\ref{sec:results:cosmology results}). 

The $p$ value, associated with the $\chi^2$-value for the best fit model is 
\begin{equation}
p = 1 - F_{\chi^2} \!\left( \chi^2, \, \nu \right),
\end{equation}
where $F_{\chi^2}$ is the cumulative distribution function of the $\chi^2$ distribution with $\nu$ degrees of freedom. In our case $\nu$ is $291$.
We find $p=0.80$, showing that there is no significant deviation of the measured data vector from the best fit model. As discussed in Appendix \ref{app: Bmodes}, we also do not detect any significant third-order $B$-modes in the KiDS-Legacy data.

\subsection{Determination of most-informing scales}

As discussed in Sect.~\ref{sc:methods:covariance}, the length of the data vector is limited by the number of simulation realisations available for the covariance estimate, restricting us to $N_\mathrm{d}=291$ entries. Since the COSEBI vector already contributes 126 modes, the selection of scales for the $\langle M_\mathrm{ap}^3\rangle$ data vector must be made with care in order to maximise cosmological information while remaining within this limit.  

To identify the most informative entries, we tested three strategies for reducing the dimensionality of the $\langle M_\mathrm{ap}^3\rangle$ data vector. A first option is to restrict the analysis to equilateral configurations with $\theta_1=\theta_2=\theta_3$. This approach was adopted in \citet{Burger2024}, who showed that it retains the majority of the available cosmological information while substantially reducing the number of independent data points. A second option is to limit the number of scales used in the lowest-redshift tomographic bins. At low redshift, the third-order signal is strongly affected by intrinsic alignments rather than by lensing, and in our modelling the IA contribution is not strongly scale dependent. In such cases, it is sufficient to constrain the IA amplitude with a single aperture radius. A third option is to create linear combinations of neighbouring tomographic bins. For example, we combine the data vector entries of bins 1 and 2 by adding them, and replace the individual values $\langle M_\mathrm{ap}^3\rangle^{(1jk)}$ and $\langle M_\mathrm{ap}^3\rangle^{(2jk)}$ both in the measured data vector and the theoretical prediction by the sum
$
    \langle M_\mathrm{ap}^3\rangle^{(1\cup 2 jk)} =
    \langle M_\mathrm{ap}^3\rangle^{(1jk)} +
    \langle M_\mathrm{ap}^3\rangle^{(2jk)}.
$
This combination reduces redundancy between bins that probe similar redshift ranges while preserving sensitivity to cosmology.  

\begin{figure*}
    \centering
    \includegraphics[width=0.49\linewidth]{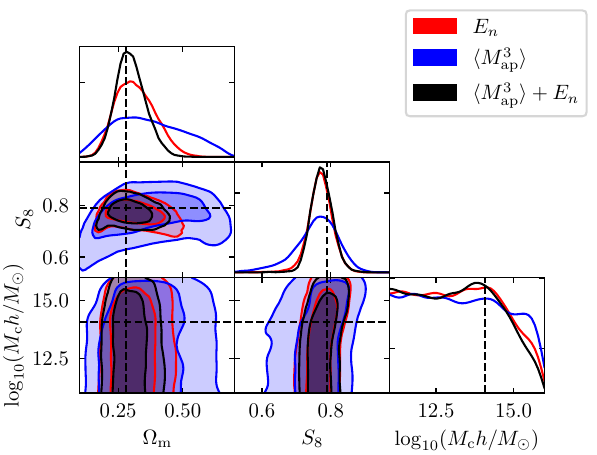}
    \includegraphics[width=0.49\linewidth]{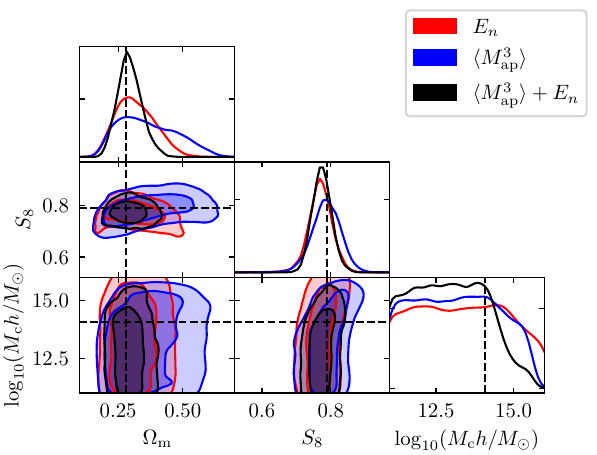}
    \caption{Parameter constraints on $\Omega_\mathrm{m}$,$S_8$, and $M_\mathrm{c}$ for a mock data vector, using either setting 2 in Table~\ref{tab: results FoM}, which uses only equi-scale aperture statistics (left) or setting 7, which includes unequal scales (right). Red are constraints from $E_n$, blue are constraints from $\langle M_\mathrm{ap}^3 \rangle$, black are joint constraints. Dark shades indicate the $1\sigma$, light shades the $2\sigma$ areas. Dashed lines denote the true parameter values of the data vector. Including the unequal scale aperture radii improves the constraints on the baryonic feedback parameter as well as the joint constraints on $\Omega_\mathrm{m}$ and $S_8$.}
    \label{fig: example constraints}
\end{figure*}

We evaluated different combinations of these strategies by applying them to a theoretically-modelled data vector and performing parameter inference with second-order statistics, third-order statistics, and the combined data vector. For each case we quantified the constraining power in the $\Omega_\mathrm{m}$–$S_8$ plane using the Figure of Merit,
\begin{equation}
    \mathrm{FoM} = \frac{1}{\sqrt{\det \mathrm{Cov}(\Omega_\mathrm{m},S_8)}},
\end{equation}
where $\mathrm{Cov}(\Omega_\mathrm{m},S_8)$ is the parameter covariance matrix. The results are listed in Table~\ref{tab: results FoM} and two example constraints shown in Fig.\ref{fig: example constraints}.

\begin{table*}
\caption{Constraining power on $\Omega_\mathrm{m}$ and $S_8$ for different choices of aperture radii and tomographic bins.}
\label{tab: results FoM}
\centering
\begin{tabular}{@{}llllllll@{}}
\toprule
\# & Aperture radii     & Tomo. bins  & Unequal scales & $N_\theta$ per bin & $N_\mathrm{d}$ & FoM $\langle M_\mathrm{ap}^3 \rangle$ & FoM $E_n$ + $\langle M_\mathrm{ap}^3\rangle$ \\ \midrule
1 & $[2', 4', 8', 16']$ & [1, 2, 3, 4, 5, 6] &  No             & [1, 4, 4, 4, 4, 4]  & 287  & 125        &   577        \\
2 & $[4', 8', 16', 32']$ & [1, 2, 3, 4, 5, 6] &  No             & [1, 4, 4, 4, 4, 4]  & 287 &  120      &  574         \\
3 & $[2', 4', 8', 16']$ & [$1 \cup 2$, 3, 4, 5, 6] &  No             & [4, 4, 4, 4, 4]  & 266 &  132      &    577       \\
4 & $[4', 8', 16', 32']$ & [$1 \cup 2$, 3, 4, 5, 6] &  No             & [4, 4, 4, 4, 4]  & 266 & 125       &   582        \\
5 & $[2', 4', 8', 16', 32']$ & [$1 \cup 2$, 3, 4, 5, 6] &  No             & [1, 4, 4, 4, 4]  & 241  &  132      &   582        \\  
6 & $[2', 4', 8', 16']$ & [$1 \cup 2$, $3 \cup 4$, $5 \cup 6$] &  Yes            & [1, 4, 4]  & 212 &   256     &  767         \\
\textbf{7} & $\mathbf{[4', 8', 16', 32']}$ & $\mathbf{[1 \cup 2, 3 \cup 4, 5 \cup 6]}$ &  \textbf{Yes }          & \textbf{[1, 4, 4]}  & \textbf{212} &  \textbf{243}      &   \textbf{789}  \\
8 & $[2', 8', 32']$ & [$1 \cup 2$, 3, 4, $5\cup6$] &  Yes            & [1, 3, 3, 3]  & 236  &   222     &    727       \\
9 & $[2', 4', 8']$ & [$1 \cup 2$, 3, 4, $5\cup6$] &  Yes            & [1, 3, 3, 3]  & 236 &   221     &   708        \\
10 & $[8', 16', 32']$ & [$1 \cup 2$, 3, 4, $5\cup6$] &  Yes            & [1, 3, 3, 3]  & 236  &  185      &   694        \\
11 & $[2', 4', 8', 16']$ & [$1 \cup 2 \cup 3$, $4 \cup 5\cup6$] &  Yes            & [4, 4]  & 206 &  211       &  707          \\
12 & $[4', 8', 16', 32']$ & [$1 \cup 2 \cup 3$, $4 \cup 5\cup6$] &  Yes            & [4, 4]  & 206 &  186      &    705       \\
13 & $[2', 4', 8', 16', 32']$ & [$1 \cup 2 \cup 3 \cup 4 \cup 5\cup6$] &  Yes            & [5]  & 161 &   125     &   694        \\
\bottomrule
\end{tabular}
\small
\textbf{Notes.} The $E_n$ data vector is chosen as in \citet{Wright2025}, i.e $n\leq 6$ and all combinations of the 6 tomographic bins. $N_\mathrm{d}$ is the length of the combined 2-point and 3-point data vector.  The FoM for $E_n$ alone is 413. Setting 7 is marked in bold, as this is our fiducial setting for all following analyses.
\end{table*}

From these results, several trends become apparent. First, the inclusion of unequal aperture scales leads to tighter constraints than restricting the analysis to equal scales. This trend is somewhat unexpected, as it runs counter to the findings of \citet{Burger2024}. A likely explanation is the adoption of the new baryonic suppression model, which \citet{Burger2025} already found to be more effectively constrained when unequal scales are included. We confirm this result (see Fig. \ref{fig: example constraints}): Without unequal scales $\langle M_\mathrm{ap}^3 \rangle$ cannot provide any information on the parameter $M_\mathrm{c}$. When including unequal scales, though, the third-order statistics become more constraining on this parameter than the second order statistics, leading also to improved $\Omega_\mathrm{m}$ and $S_8$ constraints. 

Second, the inclusion of the smallest aperture radius of 2' improves the constraining power of $\langle M_\mathrm{ap}^3\rangle$ alone, indicating that small scales carry substantial information. However, this additional information is largely degenerate with that already present in the two-point statistics, such that no significant gain is observed in the joint analysis. Finally, we find that combining tomographic bins is preferable to reducing the number of aperture radii. This suggests that the dominant information on $\Omega_\mathrm{m}$ and $S_8$, resides in the scale dependence of the aperture statistics rather than in their redshift evolution.

From this comparison we conclude that combination 7 in Table \ref{tab: results FoM} provides the optimal balance between dimensionality reduction and cosmological information. We therefore adopt this setting as the fiducial choice for all following analyses.

One might wonder why the baryonic suppression parameter $M_\mathrm{c}$ remains unconstrained in our analysis, despite forecasts by \citet{Burger2025} predicting meaningful constraints for a \textit{Euclid} DR1-like setup. We address this question in Appendix~\ref{app: bcm constraining power} and find that the combined impact of nuisance and intrinsic-alignment parameters, the more conservative scale cuts applied to KiDS, and the adopted prior on $\Omega_\mathrm{m}$ together reduce the constraining power on $M_\mathrm{c}$, without any single effect being dominant.

\subsection{Data consistency tests for third-order statistics}

Before combining the measured 2- and 3-point statistics and inferring cosmological parameters, the consistency of the third-order constraints was tested under various catalogue-level splits. The data were divided along several axes (KiDS North vs.\ South, including or excluding the smallest aperture radius scale, and including or excluding subsets of tomographic bins), and cosmological inference was repeated for each split. This analysis was carried out for all three blinded data vectors.

\begin{figure}
    \centering
    \includegraphics[width=\linewidth]{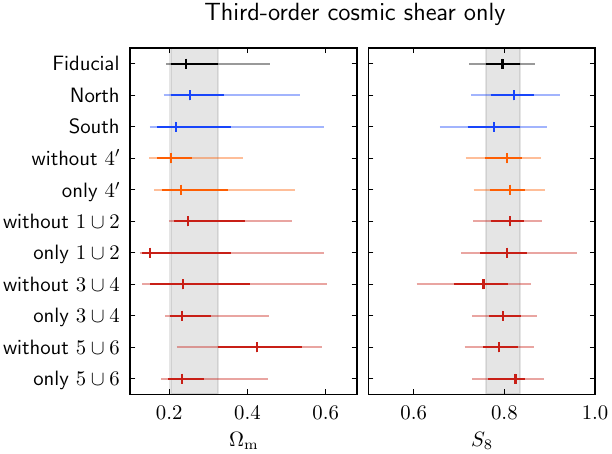}
    \caption{Constraints on $\Omega_\mathrm{m}$ and $S_8$ from $\langle M_\mathrm{ap}^3\rangle$ alone for various data splits. We mark the mode of the marginal distributions, along with the 1 and 2$\sigma$ highest posterior density intervals. High-opacity whiskers indicate the 68\% credible interval, while low-opacity whiskers indicate the 95\% credible interval. The grey band marks the 1$\sigma$ interval of the fiducial result. Note that while some splits appear to yield tighter 1D $\Omega_\mathrm{m}$ constraints, the fiducial case has the smallest area in the $\Omega_\mathrm{m}$–$S_8$ plane; narrower marginals can result from projection of elongated 2D contours.}
    \label{fig: whisker data splits}
\end{figure}

Figure~\ref{fig: whisker data splits} shows the resulting one-dimensional constraints on $\Omega_\mathrm{m}$ and $S_8$. We mark the mode of the marginal distributions, along with the 1 and 2$\sigma$ highest posterior density interval. The results from all splits are consistent with the fiducial analysis using the full data vector. No significant deviations are seen for either parameter, indicating that all splits can be described by a common set of model parameters.

In addition to this visual comparison of one-dimensional constraints, two statistical tests were performed to quantify the consistency between data splits, following the Tier~2 and Tier~3 methodology of \citet{stoelzner2025}.

For the Tier~2 test, the consistency of the parameter constraints from two data splits $A$ and $B$ was quantified using the parameter-space tension statistic $T$ from \citet{Raveri2020}. Independent MCMC chains were obtained for each split, yielding samples in the parameter space $\boldsymbol{\theta} = (\Omega_\mathrm{m}, S_8)$. From these, the posterior means $\boldsymbol{\mu}_A$, $\boldsymbol{\mu}_B$ and covariance matrices $\mathbf{C}_A$, $\mathbf{C}_B$ were estimated. The tension statistic is defined as
\begin{equation}
    T =
    (\boldsymbol{\mu}_A - \boldsymbol{\mu}_B)^\mathsf{T}
    \left(\mathbf{C}_A + \mathbf{C}_B\right)^{-1}
    (\boldsymbol{\mu}_A - \boldsymbol{\mu}_B)\,,
\end{equation}
i.e.\ the squared distance between the posterior means, normalised by the sum of the covariances. Under the null hypothesis that both splits are described by the same underlying parameters, and assuming Gaussian posteriors, $T$ follows a $\chi^2$ distribution with $N_\mathrm{par}=2$ degrees of freedom. The corresponding tension level $N_\sigma$ is defined as the one-dimensional Gaussian significance with the same two-sided tail probability as the observed $\chi^2$ value. Values $N_\sigma \lesssim 1$ indicate mutually consistent posterior means given the joint uncertainties.

For the Tier~3 test, a translated posterior distribution (TPD) approach was used following \citet{stoelzner2025}. The goal is to test whether the observed data vector of one subset, $\vec{d}_A$, is compatible with the distribution of model predictions inferred from the posterior of another subset, $B$. For each posterior sample $\boldsymbol{\theta}_B^{(i)}$ from the MCMC chain of subset $B$, the corresponding model prediction for subset $A$,
$\vec{t}_A(\boldsymbol{\theta}_B^{(i)})\,,$
is evaluated. We then draw realisations of the data vector $\vec{d}_A$ from a Gaussian distribution 
$\mathcal{N}\!\bigl(\mathbf{\mu}_A^\mathrm{sim}, \textbf{C}_A^\mathrm{sim}\bigr)$ with
\begin{align}
    \mathbf{\mu}_A^\mathrm{sim} &= \vec{t}_A(\boldsymbol{\theta}_B) + \mathbf{C}_{AB}\textbf{C}_{BB}^{-1} [\vec{d}_B - \vec{t}_B(\boldsymbol{\theta}_B)]\;,\\
    \mathbf{C}_A^\mathrm{sim} &= C_{AA} - C_{AB}C_{BB}^{-1}C_{BA}\;,
\end{align}
where $C_{AB}$ describes the cross-covariance of data sets $A$ and $B$. For each posterior sample and each simulated realisation $\vec{d}_A^{\mathrm{sim}}$, the $\chi^2$ statistic
\begin{equation}
    \chi^2[\vec{d}, \vec{t}(\boldsymbol{\theta})] 
    = \bigl[(\vec{d} - \vec{t}(\boldsymbol{\theta})\bigr])^\mathsf{T} 
      \mathbf{C}_A^{-1} 
      \bigl[(\vec{d} - \vec{t}(\boldsymbol{\theta})\bigr])
\end{equation}
is computed. In particular, for each $\boldsymbol{\theta}_B^{(i)}$ the observed and simulated discrepancies are
\begin{equation}
    \chi^2_{\mathrm{obs},i} = \chi^2\bigl[\vec{d}_A, \vec{t}_A(\boldsymbol{\theta}_B^{(i)})\bigr]\;, 
    \qquad
    \chi^2_{\mathrm{sim},i} = \chi^2\bigl[\vec{d}_{A,i}^{\mathrm{sim}}, \vec{t}_A(\boldsymbol{\theta}_B^{(i)})\bigr]\;.
\end{equation}
The TPD-based $p$-value is then defined as the fraction of TPD realisations for which the simulated data yield a higher $\chi^2$ than the observed $\vec{d}_A$. This $p$-value quantifies the probability that the data in subset $A$ are a realisation of the TPD inferred from subset $B$. Low values of $p(A|B)$ therefore indicate a potential internal inconsistency of the data.

The resulting $N_\sigma$ and $p(A|B)$ values for the different splits are listed in Table~\ref{tab: data consistency}. Overall, these tests support the consistency of the third-order data under the considered splits. No deviation reaches $1\sigma$ and no $p$-value is smaller than 0.05. In particular, no inconsistencies between small and large scales are detected, suggesting that the adopted model for baryonic effects yields mutually consistent cosmological constraints across scales.

\begin{table}
\centering
\caption{Results of consistency tests for data splits for $\langle M_\mathrm{ap}^3\rangle$}
\label{tab: data consistency}
\begin{tabular}{@{}ccccc@{}}
\toprule
A                   & B             & $N_\sigma$   & $p(A|B)$  & $p(B|A)$  \\ 
\midrule
north               & south         & 0.34         & 0.29      & 0.26      \\
without $4'$        & only $4'$     & 0.14         & 0.22      & 0.37      \\
without $1\cup2$    & only $1\cup2$ & 0.04         & 0.51      & 0.15      \\
without $3\cup4$    & only $3\cup4$ & 0.04         & 0.29      & 0.82     \\
without $5\cup6$    & only $5\cup6$ & 0.80         & 0.07      & 0.16      \\
\bottomrule
\end{tabular}
\end{table}

The largest deviation arises when including or excluding the last two tomographic bins ($p=0.07$). This could, in principle, hint at residual modelling inaccuracies or unaccounted systematic effects at higher redshifts. However, given the low significance, there is no concrete evidence that the model or systematics treatment needs to be revised for the highest tomographic bins.

\subsection{Impact of modelling choices for third-order statistics}

\begin{figure}
    \centering
    \includegraphics[width=\linewidth]{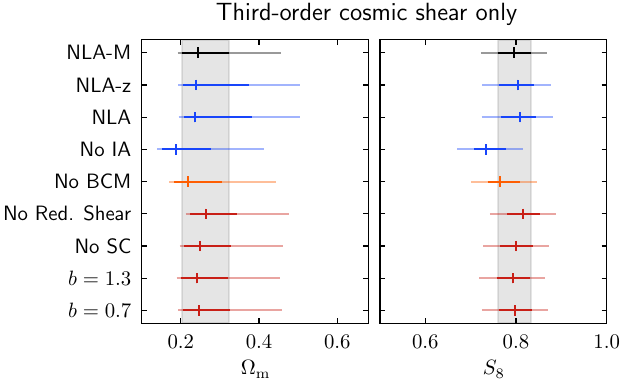}
    \caption{Inferred cosmological parameters from $\langle \Map^3 \rangle$ alone under different modelling choices. We mark the mode of the marginal distributions, along with the 1- and 2- $\sigma$ highest posterior density interval. High opacity whiskers indicate the 68\% credible interval, while low opacity whiskers indicate the 95\% credible interval. `BCM` refers to the baryonic correction model, `SC`refers to the source clustering correction.}
    \label{fig: IA consistency}
\end{figure}

We also assess the impact of our modelling choices on the inferred cosmology from $\langle M_\mathrm{ap}^3\rangle$. For this, we analyse all three blinded data vectors with all four different IA models (no IA, NLA, NLA-$z$, NLA-$M$), neglecting the impact of baryonic feedback, and neglecting the higher-order effects, namely reduced shear and source clustering. The resulting values for $\Omega_\mathrm{m}$ and $S_8$ are shown in Fig. \ref{fig: IA consistency}.

For all IA models, the best-fitting values for $\Omm$ and $S_8$ agree within $1\sigma$. Therefore, the choice of IA model is unlikely to bias the inferred cosmological constraints. However, ignoring IA completely leads to lower values of both $\Omm$ and $S_8$ compared to the fiducial analysis. While the shift is not large, it is still illustrative of the importance of including IA modelling in any cosmic shear analysis.

Neglecting the impact of baryonic feedback also affects the inferred cosmology. As expected, both the inferred $S_8$ and $\Omm$ are biased low compared to the fiducial analysis. In the case of $\Omm$ this bias amounts to $1\sigma$, while the discrepancy is smaller for $S_8$.

The higher-order effects (reduced shear and source 
clustering) both show smaller impacts on the inferred cosmology. While neglecting the reduced shear approximation leads to a $0.54 \sigma$  shifts towards higher $S_8$ values, neglecting source clustering leads to a shift of only $0.13\sigma$. We also observe that the choice of galaxy bias $b$ does not significantly impact the cosmological constraints. As expected, assuming a higher bias $b=1.3$ leads to slightly lower inferred $S_8$, while $b=0.7$ leads to higher inferred $S_8$. However, in both cases the difference is less than $0.1\sigma$. 

\subsection{Cosmological results from joint analysis}
\label{sec:results:cosmology results}

\begin{figure}
    \centering
    \includegraphics[width=\linewidth]{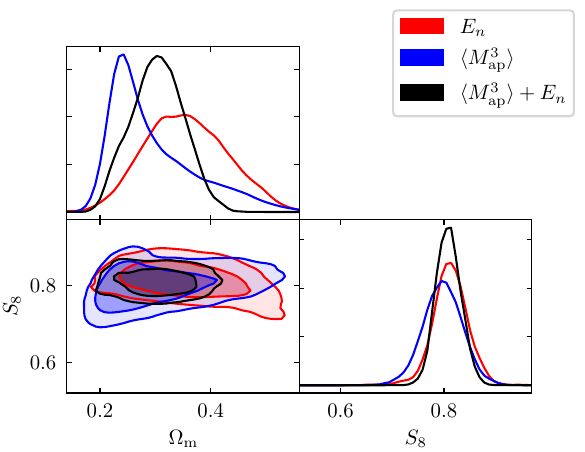}
    \caption{Constraints on $\Omm$ and $S_8$ from third-order cosmic shear statistics (red), second-order cosmic shear statistics (blue), and joint (black) for the KiDS-Legacy data set for the fiducial modelling choice}
    \label{fig: fiducial joint constraints}
\end{figure}

After assessing the data consistency and impact of modelling choices on the inferred cosmology from $\langle M_\mathrm{ap}^3 \rangle$, we combine the third-order statistics with the $E_n$ measured in \citet{Wright2025} for a joint cosmological inference. The constraints on $\Omega_\mathrm{m}$ and $S_8$ for our fiducial modelling choice are shown in Fig. \ref{fig: fiducial joint constraints}. The modes of the marginal distributions for the joint analysis are
\begin{equation}
    \Omm = 0.297^{+0.056}_{-0.040} \qquad \mathrm{and} \qquad    S_8 = 0.806^{+0.025}_{-0.023} \;,
\end{equation}
supporting that KiDS-Legacy yields constraints in both $S_8$ and $\Omega_\mathrm{m}$ that are consistent with the CMB.

The second- and third-order constraints are in excellent agreement in the $\Omega_\mathrm{m}$–-$S_8$ plane, confirming the internal consistency of the two-point and three-point cosmic shear information. The modes of the $\Omm$ distributions differ by $1.4 \sigma$, while those of the $S_8$ distributions differ by only $0.3 \sigma$. Adding $\langle M_\mathrm{ap}^3\rangle$ to the $E_n$ improves the constraints primarily on $\Omega_\mathrm{m}$, with the marginal error on this parameter reduced by about $36\%$ compared to the second-order analysis alone. The constraining power on $S_8$ also benefits from the inclusion of third-order statistics, with an uncertainty reduction of about $24\%$. Overall, the figure of merit in the $\Omega_\mathrm{m}$–$S_8$ plane increases from 429 for the $E_n$-only analysis to 895 for the combined $E_n + \langle M_\mathrm{ap}^3\rangle$ case, corresponding to an improvement by more than a factor of two. This demonstrates that the third-order information provides significant additional constraining power beyond that of the two-point statistics alone. 

We note that the marginal $S_8$ constraints from the combined $E_n + \langle M_\mathrm{ap}^3\rangle$ analysis are comparable to, rather than tighter than, the fiducial KiDS-Legacy two-point result. This reflects both the different baryonic feedback parametrization employed here, which leads to broader second-order-only constraints, and the fact that the main information gain from third-order statistics lies along a different direction in parameter space rather than aligned with the $S_8$ axis, as evidenced by the substantial improvement in the two-dimensional figure of merit.

\begin{figure}
    \centering
    \includegraphics[width=\linewidth]{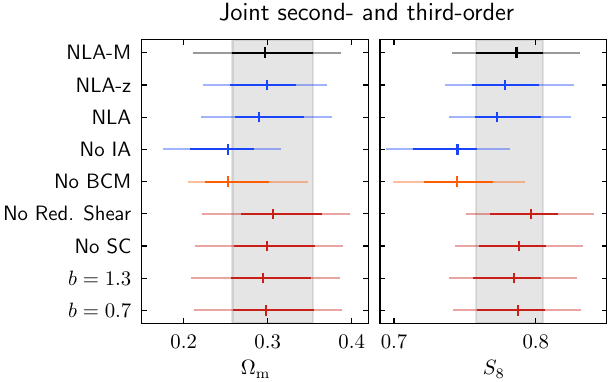}
    \caption{Comparison of joint constraints of $\langle \Map^3\rangle$ and $E_n$ under different modelling choices. The grey band shows the 1$\sigma$ constraints for the fiducial result.}
    \label{fig: model consistency}
\end{figure}

To assess the impact of the modelling assumptions, Fig.~\ref{fig: model consistency} compares the one-dimensional constraints on $\Omega_\mathrm{m}$ and $S_8$ obtained for different IA prescriptions and when ignoring the impact of baryonic feedback. All IA models considered yield mutually consistent constraints, indicating that the cosmological results are robust against variations in the IA treatment. Neglecting IA or baryonic effects, however, leads to a systematic shift towards lower values of both $\Omega_\mathrm{m}$ and $S_8$. This shows that, while the fiducial constraints are not extremely sensitive to the precise details of the IA and baryon models, including both effects is essential to avoid biased cosmological inferences.

\subsection{Comparison to other analyses}

\begin{figure}
    \centering
    \includegraphics[width=\linewidth]{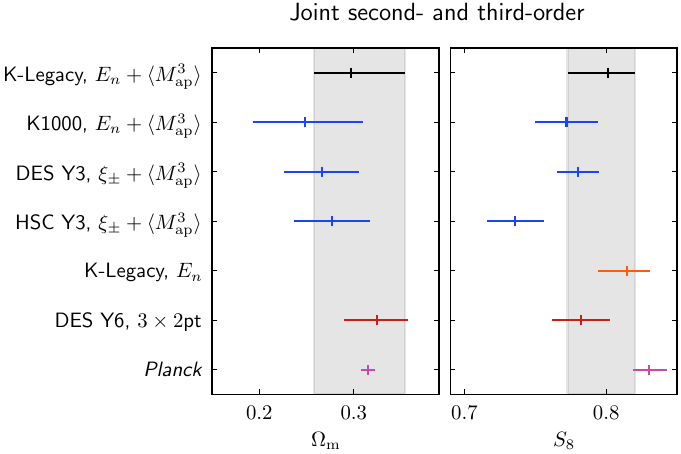}
    \caption{Comparison of the constraints from the current work (black) to other combined second- and third-order analyses (blue, from \citealp{Burger2024}, \citealp{Gomes2025}, and \citealp{Sugiyama2025}), the KiDS-Legacy second-order cosmic shear analysis (orange, from \citealp{Wright2025}), the DES Year 6 3$\times$2 analysis (red, from \citealp{DESY63x2}), and \textit{Planck} (pink, from \citealp{PlanckCosmologicalParameters}). We note that since the KiDS-Legacy second-order analysis could not constrain {\Omm} better than the prior range, no one-dimensional constraints on this parameter have been reported.}
    \label{fig: comparison external probes}
\end{figure}

We now compare the inferred constraints on $\Omega_\mathrm{m}$ and $S_8$ to results from other weak-lensing analyses in Fig.~\ref{fig: comparison external probes}. In particular, we compare the joint second- and third-order results from this work with those from KiDS-1000 \citep{Burger2024}, DES \citep{Gomes2025}, and HSC \citep{Sugiyama2025}, as well as with the second-order KiDS-Legacy cosmic shear analysis of \citet{Wright2025} and the cosmic shear, galaxyy-galaxy-lensing, and galaxy clustering (3$\times$2) analysis of DES Year 6 \citet{DESY63x2}.

The combined second- and third-order KiDS-Legacy constraints presented here are consistent within $1\sigma$ with the corresponding joint analyses from KiDS-1000 and DES in both $\Omega_\mathrm{m}$ and $S_8$. They are also broadly consistent with the HSC joint analysis, although the HSC result prefers a lower value of $S_8$. This difference remains below the $2\sigma$ level but is noticeable. A possible explanation is the absence of an explicit baryonic feedback model in the HSC analysis, which may lead to a residual suppression of small-scale power being absorbed into a lower inferred $S_8$.

In terms of constraining power, the joint KiDS-Legacy results are comparable to those from other second and third-order analyses. The DES Y3 combination achieves slightly tighter error bars, which is attributable to its larger survey area. Relative to the KiDS-1000 second and third-order analysis, the present constraints favour a higher value of $S_8$, in line with the shift observed between the KiDS-1000 and KiDS-Legacy two-point analyses.

A direct comparison to the fiducial KiDS-Legacy second-order cosmic shear analysis \citep{Wright2025} shows excellent agreement in $S_8$, with the joint $E_n + \langle M_\mathrm{ap}^3\rangle$ constraints lying well within $1\sigma$ of the two-point result. %The second-order KiDS-Legacy analysis does not constrain $\Omega_\mathrm{m}$ more tightly than the prior, so a one-dimensional comparison for this parameter is not available.
Our results also agree with the most recent DES results \citep{DESY63x2}. These were obtained using the combination of second-order cosmic shear, galaxy-galaxy lensing, and galaxy clustering, i.e., a ``3$\times$2-point'' analysis.

Finally, the joint constraints obtained from KiDS-Legacy are fully consistent with the \textit{Planck} \citep{PlanckCosmologicalParameters} results for both $\Omega_\mathrm{m}$ and $S_8$ within $1\sigma$. There is therefore no evidence for an $S_8$ tension in the KiDS-Legacy data for either second-order or the combined second- and third-order cosmic shear statistics under the adopted modelling framework.

\section{\label{sc:Discussion} Discussion}

This work presents a joint analysis of second- and third-order cosmic shear statistics in the final data release from the Kilo-Degree Survey (KiDS), the KiDS-Legacy data set.  COSEBIs are used as second-order summary statistics and the third-order aperture mass $\langle M_\mathrm{ap}^3\rangle$ as the higher-order probe, modelled with a redshift-resolved bispectrum emulator, an IA model with redshift- and mass-dependence, a physically motivated baryon correction model, and corrections for reduced shear and source clustering.

A key technical result is the optimisation of the third-order data vector under the constraints imposed by the covariance estimation. By jointly selecting aperture radii and tomographic combinations, we show that including non-equal aperture scales significantly enhances the constraining power of $\langle M_\mathrm{ap}^3\rangle$. Unequal scales improve sensitivity in particular to the baryonic suppression parameter $M_c$ and, through this, to $\Omega_\mathrm{m}$ and $S_8$. This contrasts with earlier analyses \citep{Burger2024} based on simpler baryon treatments and highlights that the information content of third-order statistics depends sensitively on the interplay between scale selection and baryonic modelling. At the same time, the analysis confirms that most of the cosmological constraining power of $\langle M_\mathrm{ap}^3\rangle$ arises from its scale dependence rather than from fine tomographic resolution, motivating the adopted compression of tomographic bins.

Before combining second- and third-order statistics, the internal consistency of the measured $\langle M_\mathrm{ap}^3\rangle$ was tested under a series of catalogue-level splits. These tests probe North–South differences, the impact of including the smallest aperture scale, and the contribution of different tomographic subsets. Both the Tier~2 parameter-space tension statistic and the Tier~3 translated posterior tests indicate that the third-order constraints are stable under these splits. No configuration exhibits a deviation at the $1\sigma$ level or a TPD $p$-value below 0.05, and there is no evidence for inconsistencies between small and large scales. This finding supports the high level of consistency found for the second-order KiDS-Legacy measurements \citep{stoelzner2025}.

The joint KiDS-Legacy $E_n + \langle M_\mathrm{ap}^3\rangle$ analysis yields cosmological constraints that are fully consistent with the two-point KiDS-Legacy results of \citet{Wright2025}, while providing additional information that is not accessible to second-order statistics alone. In particular, the third-order measurements tighten the constraints on $\Omega_\mathrm{m}$ by about $36\%$ and on $S_8$ by about $24\%$ relative to the $E_n$-only analysis, resulting in a more than twofold increase of the figure of merit in the $\Omega_\mathrm{m}$–-$S_8$ plane. This behaviour is in line with the expectation that the non-Gaussian information contained in third-order shear statistics helps to break the degeneracy between $\Omega_\mathrm{m}$ and $\sigma_8$ inherent to two-point functions, and thereby extracts additional cosmological information from the same survey data.

The inferred values of $\Omega_\mathrm{m}$ and $S_8$ from the joint analysis are in agreement with those obtained from earlier combined second- and third-order studies in KiDS-1000 \citep{Burger2024}, DES \citep{Gomes2025}, and HSC \citep{Sugiyama2025} within the statistical uncertainties. The KiDS-Legacy results lie close to the DES Y3 and KiDS-1000 constraints in both parameters and remain broadly consistent with the HSC analysis, which prefers a lower $S_8$ but still agrees at better than the $2\sigma$ level. Differences between the various surveys are likely driven by a combination of survey depth, area, and modelling choices, including the treatment (or omission) of baryonic feedback. Compared to KiDS-1000, the KiDS-Legacy second and third-order constraints favour a slightly higher $S_8$, paralleling the shift seen between the corresponding two-point analyses.

Relative to the KiDS-1000 2+3-point analysis, our joint constraints favour a higher $S_8$ value that is more compatible with \textit{Planck}. When combining $E_n$ and $\langle M_\mathrm{ap}^3\rangle$, the inferred $(\Omega_\mathrm{m}, S_8)$ are fully consistent with the \textit{Planck} cosmological parameters at the $1\sigma$ level. Within the modelling framework adopted here, including IA, baryonic feedback, and higher-order lensing corrections, there is therefore no evidence for an $S_8$ tension in the KiDS-Legacy data.

%At the same time, the comparison of error bars demonstrates that KiDS-Legacy is approaching, but not yet matching, the constraining power anticipated for Stage-IV surveys. 
The combined KiDS-Legacy constraints are competitive with other current 2+3-point analyses and only modestly weaker than DES Y3, despite the smaller survey area, owing to the improved data quality and modelling. However, the baryonic suppression parameters remain essentially unconstrained. As discussed in Appendix~\ref{app: bcm constraining power}, this is consistent with expectations once the full set of nuisance parameters and modelling uncertainties is taken into account. Forecasts for Stage-IV surveys, however, indicate that constraints on the baryonic parameters from 2+3-point cosmic shear might be possible \citep{Burger2025}.

Looking ahead, the methods developed here are directly applicable to upcoming surveys such as \textit{Euclid} and LSST. The redshift-resolved bispectrum emulator and the flexible baryon correction model are designed to operate with arbitrary source redshift distributions and can thus be used for a wide range of survey configurations. The KiDS-Legacy analysis demonstrates that robust measurements of third-order cosmic shear are already feasible on current wide-field data and that a consistent joint modelling of second- and third-order statistics, including IA, baryons, reduced shear, and source clustering, can be achieved in practice. With the increased area, depth, and statistical power of Stage-IV surveys, these methods will enable substantially tighter constraints on late-time structure growth and, in particular, on the origin of any residual tension between weak lensing and early-Universe probes.

%
% Add the acknowledgement using the achnowledgements environment.
% Do not use \acknowledgement{....} as this affects the formatting
% of the references.
%

\begin{acknowledgements}

Based on observations made with ESO Telescopes at the La Silla 
Paranal Observatory under programme IDs 179.A-2004, 177.A-3016, 177.A-3017, 
177.A-3018, 298.A-5015. 

LL thanks D.A.L. for provided time and assistance. LL is supported by the Austrian Science Fund (FWF) [ESP 357-N]. This research was funded in part by the Austrian Science Fund (FWF) 10.55776/F101300. We acknowledge financial support from the Canadian Space Agency (Grant No. 23EXPROSS1) and the Waterloo Centre for Astrophysics. JHD acknowledges support from an STFC Ernest Rutherford Fellowship (project reference ST/S004858/1) for the earlier part of this work. LP acknowledges support from the DLR grant 50QE2302.

MA is supported by the UK Science and Technology Facilities Council (STFC) under grant number ST/Y002652/1 and the Royal Society under grant numbers RGSR2222268 and ICAR1231094. MB is supported by the Polish National Science Center through grant no. 2020/38/E/ST9/00395. CH acknowledges support from the Max Planck Society and the Alexander von Humboldt Foundation in the framework of the Max Planck-Humboldt Research Award endowed by the Federal Ministry of Education and Research, and the UK Science and Technology Facilities Council (STFC) under grant ST/V000594/1. H. Hildebrandt is supported by a DFG Heisenberg grant (Hi 1495/5-1), the DFG Collaborative Research Center SFB1491, an ERC Consolidator Grant (No. 770935), and the DLR project 50QE2305. HHo acknowledges support from the European Research Council (ERC) under the European Union’s Horizon 2020 research and innovation program with Grant agreement No. 101053992. BJ acknowledges support by the ERC-selected UKRI Frontier Research Grant EP/Y03015X/1 and by STFC Consolidated Grant ST/V000780/1. SJ acknowledges the Ramón y Cajal Fellowship (RYC2022-036431-I) from the Spanish Ministry of Science and the Dennis Sciama Fellowship at the University of Portsmouth. SSL has received funding from the programme ``Netzwerke 2021'', an initiative of the Ministry of Culture and Science of the State of Northrhine Westphalia. LM acknowledges the financial contribution from the grant PRIN-MUR 2022 20227RNLY3 “The concordance cosmological model: stress-tests with galaxy clusters” supported by Next Generation EU and from the grant ASI n. 2024-10-HH.0 ``Attività scientifiche per la missione Euclid – fase E''. RR is partially supported by an ERC Consolidator Grant (No. 770935). BS acknowledges support from the Max Planck Society and the Alexander von Humboldt Foundation in the framework of the Max Planck-Humboldt Research Award endowed by the Federal Ministry of Education and Research. AHW is supported by the Deutsches Zentrum für Luft- und Raumfahrt (DLR), under project 50QE2305, made possible by the Bundesministerium für Wirtschaft und Klimaschutz, and acknowledges funding from the German Science Foundation DFG, via the Collaborative Research Center SFB1491 ``Cosmic Interacting Matters - From Source to Signal''. YZ acknowledges the studentship from the UK Science and Technology Facilities Council (STFC).

We acknowledge the use of the following software: \verb|NumPy| \citep{numpy}, \verb|SciPy| \citep{scipy}, \verb|treecorr| \citep{Jarvis2004}, \verb|HEALPix| \citep{healpix}, \verb|matplotlib| \citep{matplotlib}, \verb|niceplots| \citep{niceplots}. 

\textit{Author Contributions:} 
All authors contributed to the development and writing of this paper. The
authorship list is given in two groups: the lead authors, 
followed an alphabetical group. This group covers those who 
have either made a significant contribution to the preparation of data products or to the 
scientific analyses of KiDS-Legacy.
\end{acknowledgements}

\bibliography{Euclid, my} 

\begin{appendix}
\onecolumn %If you don't want single column for the Appendix, please
             %comment this out
  %\commLaila{Appendix limited to 8 pages!}

\section{Emulator accuracy and hyperparameters}
\label{app: emulators}

As mentioned in Sect. \ref{sc:Methods}, we optimise the hyperparameters of our emulators using Bayesian optimisation. We list in Table \ref{tab: emulator architecture} the most important hyperparameters of the optimal configuration we found, namely the number of layers, the nodes per layer, and the activation function. In particular the choice of activation function was found to be significant; we found that a gelu-activation function, given as
\begin{equation}
\mathrm{gelu}(x)
= \frac{1}{2} x \left[1 + \tanh\!\left(\sqrt{\frac{2}{\pi}} \left(x + 0.044715\,x^{3}\right)\right)\right] \, ,
\end{equation}
leads to much better accuracy than other standard choices like the relu or sigmoid functions. Additional hyperparameters, such as the learning rate or the batch size did not impact the emulator accuracy significantly. 

\begin{table}
\centering
\caption{Network architecture for emulators.}
\label{tab: emulator architecture}
\begin{tabular}{@{}llll@{}}
\toprule
                                              & Number layers & Nodes per layer & Activation\\ \midrule
$\langle M_\mathrm{ap}^3 \rangle_z (\theta)$ & 6             & 128             & gelu                            \\
$E_n$ (NLA-$M$)                               & 5             & 64              & gelu                         \\
$E_n$ (NLA-$z$)                               & 6             & 32              & gelu                        \\
$E_n$ (NLA)                                 & 5             & 32              & gelu                        \\ \bottomrule
\end{tabular}
\end{table}

\begin{figure}
\begin{minipage}{0.48\linewidth}
    %\centering
    \includegraphics[width=\linewidth]{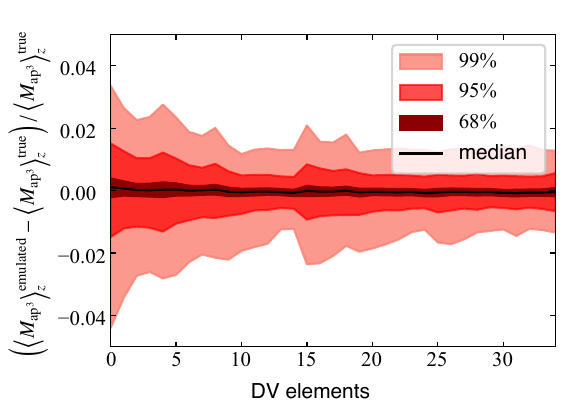}
    \caption{Accuracy of emulator for $\langle \Map^3\rangle_z$. The datavector (DV) contains the values for unique combinations of $(\theta_1, \theta_2, \theta_3)$ with $\theta_i \in [4', 8', 16', 32']$, ordered from smallest to largest scales. The peaks in the accuracy correspond to the transition from one aperture scale combination including only large scales to a combination including also a smaller scale.}
    \label{fig: Map3 emu accuracy}
\end{minipage} \hfill
\begin{minipage}{0.48\linewidth}
    %\centering
    \includegraphics[width=\linewidth]{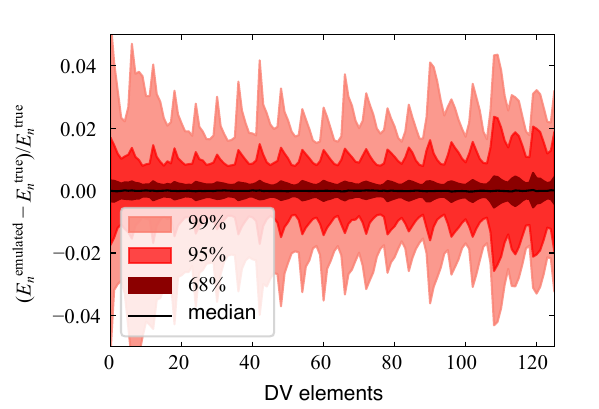}
    \caption{Accuracy of $E_n$ emulator for NLA-$M$ model. The datavector (DV) contains the first six $E_n$ for each tomographic bin combination and is sorted by tomographic bin from lowest to highest. The peaks in the accuracy correspond to the transition between different tomographic bins and thus going from $n=6$ to $n=1$.}
    \label{fig: En emu accuracy}
\end{minipage}
\end{figure}

We show in Figs \ref{fig: Map3 emu accuracy} and \ref{fig: En emu accuracy} the accuracy of our optimal emulators for the fiducial models for the third-order aperture statistics and the $E_n$. The accuracy of the emulators increases for larger aperture radii, as well as for higher $n$ modes of the COSEBIs. In both cases we find a per cent-level accuracy.

\section{Impact of baryons on third-order aperture statistics in KiDS-Legacy setup}
\label{app: impact baryons}
\begin{figure}
    \centering
    \includegraphics[width=\linewidth]{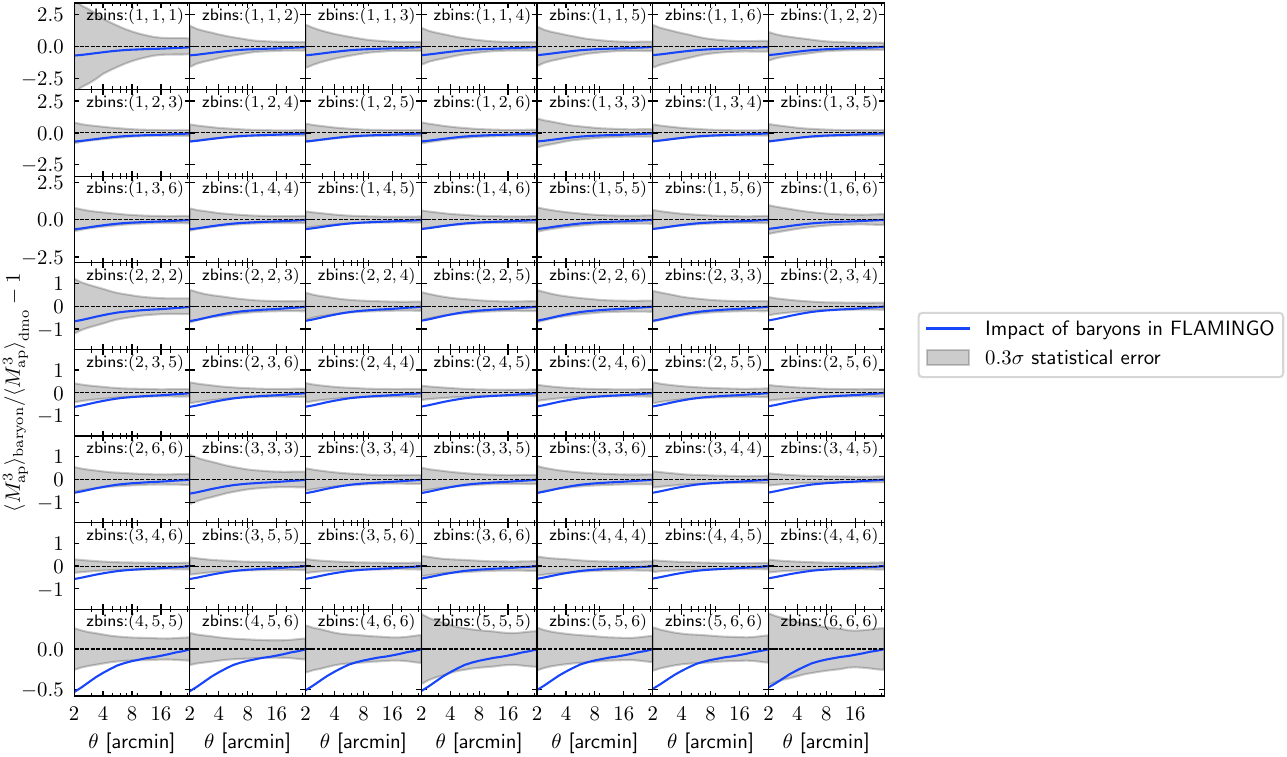}
    \caption{Relative difference of $\langle M_\mathrm{ap}^3\rangle(\theta, \theta, \theta)$ with and without the impact of baryonic effects. The baryonic effects are modelled by the average strength found in the FLAMINGO simulations by \citet{Burger2025}. The grey area shows 0.3 of the statistical uncertainty, estimated from the covariance matrix estimate. Each panel shows a different tomographic bin combination.}
    \label{fig: impact baryons}
\end{figure}

To demonstrate the necessity of including a model for baryonic effects in our analysis, we show in Fig.\ref{fig: impact baryons} the relative difference between a theoretical data vector for $\langle M_\mathrm{ap}^3\rangle$ with and without baryonic effects, along with 0.3 times the statistical uncertainty on the KiDS-Legacy measurements. The strength of the baryonic effects mimics the FLAMINGO simulations and was obtained in \citet{Burger2025}. 

A systematic effect that reaches a third of the statistical uncertainty can bias inference results in cosmological analyses if not taken into account \citep{Krause2021, Deshpande-EP28}. If multiple systematic effects compound, even lower thresholds might already cause systematic biases. We observe that for some tomographic bins the threshold of 0.3 $\sigma$ is already reached for aperture statistics as large as \ang{;10;}. Consequently, discarding scales for which baryons are negligible would result in a loss of large parts of our data vector. 

\section{Constraining power for baryonic parameters}
\label{app: bcm constraining power}
\begin{figure}
    \centering
    \includegraphics[width=0.49\linewidth]{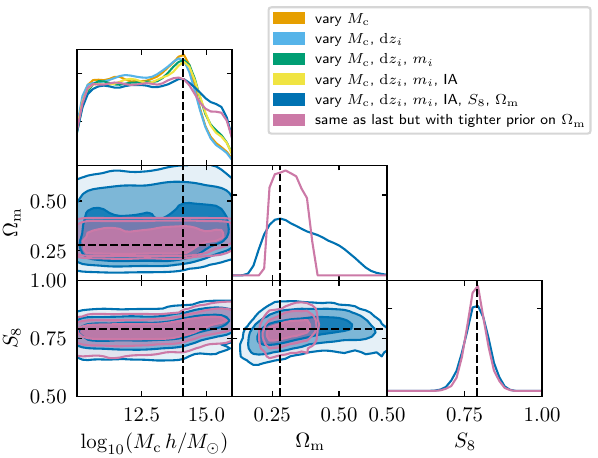}
    \includegraphics[width=0.49\linewidth]{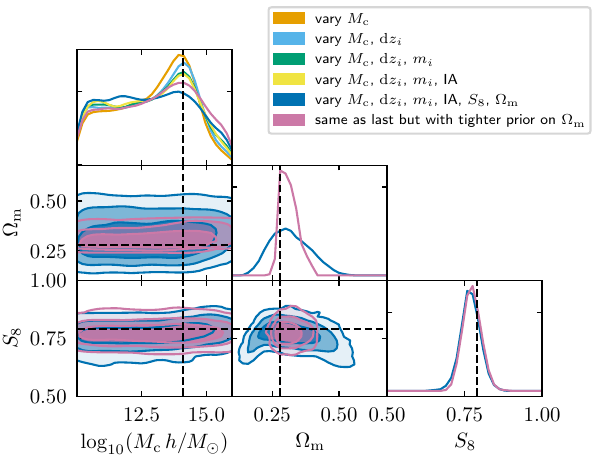}
    \caption{Constraints on mock data vector of $\langle M_\mathrm{ap}^3\rangle$ (left) and of $E_n$ (right). From top to bottom in the legend we increase the number of free parameters in the analysis. The fiducial analysis corresponds to the dark blue case.}
    \label{fig: bcm constraining power}
\end{figure}

We investigate why the KiDS Legacy data in our setup does not provide meaningful constraints on the baryonic suppression parameters, even though forecasts for a \textit{Euclid} DR1-like analysis \citep{Burger2025} predict substantial constraining power. Figure~\ref{fig: bcm constraining power} illustrates constraints obtained from theoretical second- and third-order data vectors with increasing number of free parameters. In both cases, the constraining power on $M_\mathrm{c}$ decreases steadily as additional effects are included. This indicates that there is no single dominant source of degeneracy, but rather that the KiDS data lack the statistical power to constrain all relevant effects simultaneously.

We also find that the prior on $\Omega_\mathrm{m}$ plays a role: removing very large values of $\Omega_\mathrm{m}$ (above $\sim 0.4$) through the prior also reduces the probability density for correspondingly high values of $M_\mathrm{c}$. As \citet{Burger2025} used tighter $\Omega_\mathrm{m}$ priors, their results might be affected. Furthermore, the constraining power of the $E_n$ alone is weaker than reported in \citet{Burger2025}, which can be attributed to the different scale ranges used. While their analysis of $\xi_\kappa$ extends down to 0.5 arcmin, our $E_n$ start at $2'$, limiting sensitivity to baryonic effects at small scales. Taken together, these findings suggest that the absence of strong baryonic constraints from the KiDS Legacy data is expected.

\section{\texorpdfstring{$B$}{B}-modes}
\label{app: Bmodes}

\begin{figure*}
    \centering
    \includegraphics[width=\linewidth]{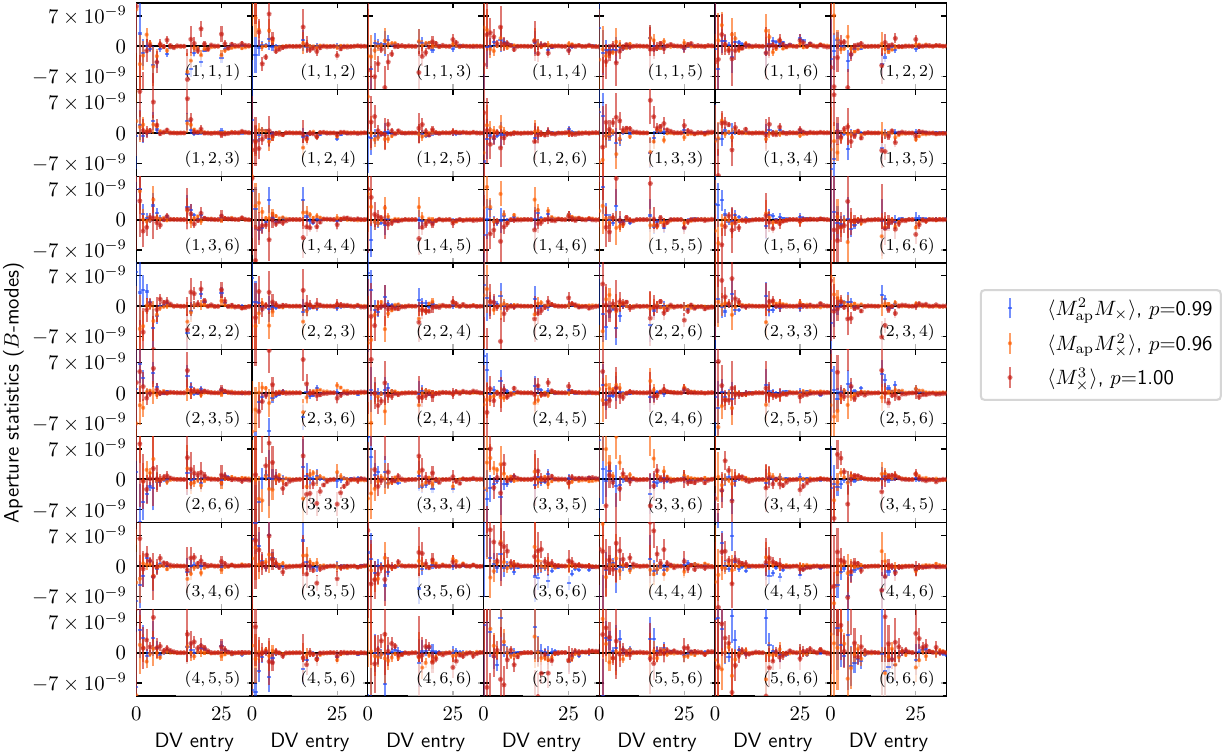}
    \caption{Measured $\langle M_\times M_\mathrm{ap}^2 \rangle$,  $\langle M_\times^2 M_\mathrm{ap} \rangle$, and $\langle M_\times^3 \rangle$, together with their significance expressed as $p$-values.}
    \label{fig: measurement Bmodes}
\end{figure*}

In Fig.~\ref{fig: measurement Bmodes} we show the $B$-modes $\langle M_\times^3 \rangle$, $\langle M_\times^2 M_\mathrm{ap} \rangle$, and $\langle M_\times M_\mathrm{ap}^2 \rangle$. In the absence of systematic effects, these should be consistent with zero. We estimate the significance of the $B$-modes using
\begin{equation}
p = 1 - F_{\chi^2} \!\left( \chi^2_\mathrm{B}, \, \nu \right),
\end{equation}
with
\begin{equation}
\chi^2_\mathrm{B} 
= \vec{d}_\mathrm{B}^\mathrm{T} \, \mathbf{C}_\mathrm{B}^{-1} \, \vec{d}_\mathrm{B} \,.
\end{equation}
We find for $\langle M_\times^3 \rangle$ $p=1.0$, 
for $\langle M_\times^2 M_\mathrm{ap} \rangle$ $p=0.96$, 
and for $\langle M_\times M_\mathrm{ap}^2 \rangle$ $p=0.99$. 
Consequently, no significant $B$-modes are present in the data.

% \section{Additional posterior constraints}
% \commLaila{For fiducial joint analysis show here all parameter constraints}

\end{appendix}

\label{LastPage}
\end{document}